\documentclass[journal]{IEEEtran}
\usepackage{float}
\usepackage{comment}
\usepackage{siunitx}
\usepackage{xcolor}
\usepackage[hyphens]{url}
\usepackage{hyperref}
\usepackage{dsfont}
\usepackage{bm}
\usepackage{amsmath,amssymb,amsfonts}
\usepackage{algorithmic}
\usepackage{algorithm}
\usepackage{array}
\usepackage[caption=false,font=normalsize,labelfont=sf,textfont=sf]{subfig}
\usepackage{textcomp}
\usepackage{stfloats}
\usepackage{verbatim}
\usepackage{graphicx}

\usepackage{cite}
\usepackage{tikz}
\usetikzlibrary{positioning, shapes.geometric, arrows.meta, backgrounds}
\newenvironment{smallalign}[1][\small]
{%
    \par\nopagebreak
    #1%
    \leavevmode\vspace*{-\baselineskip}%
    \align
}
{%
    \endalign
    \ignorespacesafterend
}

\newenvironment{smallalign*}[1][\small]
{%
    \par\nopagebreak
    #1%
    \leavevmode\vspace*{-\baselineskip}%
    \csname align*\endcsname
}
{%
    \csname endalign*\endcsname
    \ignorespacesafterend
}

\newif\ifSMonly

\begin{document}

\title{PHAST-Net: Attention-Guided, Physics-Informed Network for Estimation of Ideal Time--Frequency Representations}

\author{James M. Cozens,~\IEEEmembership{Member,~IEEE,} Simon J. Godsill,~\IEEEmembership{Fellow,~IEEE} \vspace{-1.25mm}
\thanks{This work has been submitted to the IEEE for possible publication. Copyright may be transferred without notice, after which this version may no longer be accessible.}
\thanks{J. Cozens and S. Godsill are with the Engineering Department, University of Cambridge, CB2 1PZ Cambridge, U.K., E-mail: (jmc257@cam.ac.uk; sjg30@cam.ac.uk).
\textit{(Corresponding author: James M. Cozens)}}% 
\thanks{The work of James M. Cozens was supported by EPSRC Doctoral Training Partnership (DTP) under Grant 2738835}
\thanks{Potential conflict of interest: J. M. Cozens is a named inventor and patentee of pending UK patent application GB2614708.2, relating to aspects of the technology described in this article.}
}

\maketitle

\begin{abstract}
We introduce PHAST-Net, an attention-guided, physics-informed network for unified estimation of Ideal Time--Frequency Representations (ITFRs), spanning spectral, tempo-based, metrical, and harmonic representations such as Spectrograms, Tempograms, and Metrograms. PHAST-Net learns a mapping from a constellation of wavelet transforms, specifically the proposed Continuous Log-frequency Adaptive Wavelet Transform (CLAWT), to high-resolution, cross-term-suppressed time-frequency representations. The proposed constellation of CLAWTs is strategically selected through Wigner--Ville kernel analysis to provide orientation-diverse coverage in a log–frequency time–frequency plane tailored to harmonic signal structure. PHAST-Net further incorporates a proposed physics-informed auxiliary reprojection loss designed to match the CLAWT-domain reprojections of the predicted and target ITFRs using the corresponding CLAWT kernels during training. This auxiliary objective promotes transform and energy-distribution consistency, mitigates target sparsity, and regularises optimisation. Attention layers further promote effective cross-term suppression across the input constellation. The log-frequency formulation also enables Harmonic PHAST-Net, which estimates a Harmonic ITFR that isolates fundamental structure, supporting robust fundamental-only representations for speech and music, such as derived fundamental Tempograms and Metrograms. We further introduce Spline-PHAST-Net, which parameterises detected and associated time–frequency ridges as continuous spline trajectories, enabling arbitrary-grid re-rendering and signal reconstruction. Trained on a procedurally generated dataset, PHAST-Net demonstrates improved accuracy over representative time--frequency analysis methods, providing a unified framework for high-resolution, cross-term-robust analysis of speech, music, and broader nonstationary signals.

\end{abstract}


\vspace{-2mm} \section{Introduction}

\vspace{-2mm} \subsection{Motivation}

\IEEEPARstart{T}{ime-frequency} (T--F) analysis serves as a fundamental tool for studying nonstationary processes, offering interpretable structure for complex signals whose spectral content evolves over time \cite{Boashash2016, applications}. In speech, music, and wider audio processing, T--F representations support transcription, signal decomposition, classification, restoration, segmentation, and visualisation \cite{Cozens, classification, transcription}, increasingly functioning as both inputs and targets within deep learning pipelines, including modern generative models \cite{generation}. Beyond audio, T--F methods are routinely employed for EEG and ECG based neurological and cardiac assessment \cite{Zhang2023, Alazrai2018, Hussein2018, Escriva2018}, vibration and acoustic monitoring for fault diagnosis and predictive maintenance \cite{Feng2013}, analysis of financial time series for volatility and anomaly structure \cite{finance1, finance2}, and time-localised spectral characterisation of transient behaviour in gravitational-wave observatories \cite{Abbott2020LIGONoiseGuide}. For harmonic-rich audio signals, fundamental-focused representations improve speech and music pitch-centric inference and visualisation, in addition to enhancing spectral representations such as Tempograms and Metrograms \cite{transcription, LinSuWu2018DeShapeSST, Cozens}. Harmonic-suppressed T--F representations are also important in applications such as power systems (synchrophasor and PLL phase–frequency tracking), rotating machinery (tacholess order tracking), vibroseis seismic processing (harmonic-noise removal), and biomedical examinations (ECG and EEG) \cite{XuChattertonPennacchiLiu2020TOT, LiuEtAl2022VibroseisHarmonicNoise, LinSuWu2018DeShapeSST}.

\vspace{-3mm} \subsection{Prior Work}

\IEEEpubidadjcol A core objective in T--F analysis is to obtain representations that remain sharply concentrated along physically meaningful T--F signal trajectories whilst suppressing cross-term interference and retaining practical computational tractability \cite{Boashash1992,Cohen1989,Flandrin1999}. Bilinear representations are especially relevant in this regard, given their ability to attain high auto-term concentration. In particular, the Wigner--Ville distribution (WVD) provides optimal concentration for linear frequency-modulated T--F trajectories, whilst Polynomial Wigner--Ville distributions (PWVDs) extend this optimality to higher-order frequency-modulated signals \cite{PWVD}. However, the direct application of both the WVD and PWVD to multicomponent signals is limited by strong cross-term interference \cite{wvd,wvdinter,Boashash1992,Flandrin1999,Cohen1989}. In addition, the increased interference complexity, computational burden, and reduced analytical convenience of the PWVD often make WVD-based methods more viable in practice for T--F analysis \cite{PWVD,Boashash2016,Cohen1989}. In particular, WVD-derived bilinear representations, including spectrograms, Continuous Wavelet Transform (CWT) energy, and fractional-wavelet energy formulations, as well as adaptive optimal-kernel methods, provide mechanisms for suppressing WVD cross-term interference, typically at the expense of reduced auto-term concentration \cite{Cohen1989,wavelet,Jones,jones2,Mohammadi,Awal2017,awal2,Boashash2016}. Likewise, approaches such as spectral reassignment, synchrosqueezing, deconvolutional methods, and multiresolution techniques improve T--F concentration, but typically incur the same underlying trade-off between interference suppression, localisation, and computational tractability \cite{reassignment,Flandrin2002,Thakur2013,yu2017set,stankovic1994smethod,ref:1,ref:2,deconv_TF_multiple}. Decomposition-based alternatives, including Empirical Mode Decomposition and Variational Mode Decomposition, as well as dictionary-based methods such as matching pursuit, including harmonic formulations for audio \cite{Gribonval2003}, provide sparse component extraction frameworks. However, they typically require mode- or dictionary-specific assumptions, remain sensitive to mode mixing and noise, and can become computationally burdensome for dense multicomponent signals \cite{Huang1998,VMD,Mallat}. Motivated by these limitations, the RIFT algorithm \cite{cozens2025rift} introduced a model-based framework for jointly suppressing cross-term interference and sharpening T--F representations via iterative deconvolution over a strategically chosen constellation of fractional wavelet transforms spanning multiple orientations and standard deviations. 

More recently, deep-learning-based methods have emerged as a promising approach to T--F analysis. Direct learned reconstruction of cross-term-suppressed idealised representations has been explored through CNN-based and data-driven cross-term-free T--F estimation, semi-supervised WVD refinement, learned structured-sparsity reconstruction, and GAN-based WVD enhancement \cite{Zhang2022CTFreeCNN,Liu2023WVDNet,Jiang2020UISTA,Jiang2022DHTFD,Quan2024WVDGAN,Yang2022SparseTFNet}. Furthermore, broader learned T--F models, including TFA-Net, QTFN, and adaptive multi-scale TF-net, demonstrate the potential of neural inference to further enhance concentration and suppress cross-term interference \cite{Pan2023TFANet,Chen2024AMTFN,Chen2024QTFN}. However, existing learned approaches have largely focused on enhancing fixed T--F representations or directly learning analysis operators, rather than estimating ideal representations under explicit transform-family consistency constraints. Additionally, several recent learned methods continue to report challenges for dense multicomponent signals with intersecting trajectories, particularly in noisy conditions. Moreover, although classical logarithmic-frequency methods such as the constant-Q transform are well established \cite{Brown1991CQT}, and refinement methods have also been explored in filter-bank and log-frequency formulations \cite{Fenet2017ReassignedCQT, Holighaus2016FilterBankReassignment}, comparatively few works address unified concentration enhancement and interference suppression directly in a log-frequency framework tailored to harmonic signal structure, particularly in audio applications.

Harmonic-suppressed T--F representations present a distinct challenge in T--F analysis due to the time- and frequency-varying nature of overtone structure encountered in many real-world signals \cite{transcription,XuChattertonPennacchiLiu2020TOT,LiuEtAl2022VibroseisHarmonicNoise,LinSuWu2018DeShapeSST,Cozens}. Fundamental-extracting approaches have been proposed, including harmonic matching-pursuit decompositions for audio \cite{Gribonval2003}. The de-shape synchrosqueezing method similarly reduces wave-shape and harmonic influence to more directly reflect underlying fundamental behaviour \cite{LinSuWu2018DeShapeSST}, whilst methods such as the YIN algorithm and variable-rate particle-filter approaches provide frameworks for monophonic or speech-oriented fundamental tracking \cite{deCheveigne2002YIN, Zhang2016}. However, synchrosqueezing can still exhibit cross-term interference, monophonic trackers do not generalise to polyphonic T--F representations, and fundamental--harmonic ambiguity persists \cite{LinSuWu2018DeShapeSST,transcription}.  

Tempograms are time--tempo representations that capture how a signal’s periodicity evolves over time, making them especially useful for music processing tasks such as beat tracking and tempo estimation \cite{ref:tempogram1, ref:tempogram2, Cozens}. A Tempogram is typically computed by applying T--F analysis to an onset detection (novelty) function derived from the audio, most often via the Short-Time Fourier Transform (yielding a Fourier Tempogram) or via Autocorrelation (yielding an Autocorrelation Tempogram) \cite{ref:tempogram1, ref:tempogram2, Cozens}. However, Fourier-based approaches produce rhythmic harmonics, whereas Autocorrelation-based Tempograms produce sub-harmonics, thereby reducing interpretability. Variants such as cyclic Tempograms \cite{ref:tempogram1} and methods that combine the properties of the Fourier- and Autocorrelation-based representations \cite{ref:tempogram2, Cozens, Gui2018TempogramMP} aid in suppressing tempo-harmonic ambiguity, with enhanced concentration achieved through a Matching Pursuit-based construction \cite{Gui2018TempogramMP}. However, the time--tempo representations still exhibit cross-term interference, reduced concentration, and mode-mixing, analogous to those encountered in classical T--F analysis. More recently, the Metrogram was introduced as a transform that extracts time-varying rhythmic and metric periodicities from Tempogram representations, supporting dynamic time-signature inference, tempo analysis, and beat tracking \cite{Cozens}. As a result of the construction, any harmonic ambiguity or cross-term interference present in the underlying tempogram representation can be misinterpreted by the Metrogram as metric information; thus, a sharpened, cross-term- and harmonic-suppressed input representation is crucial for the Metrogram's performance.

\vspace{-2mm}\subsection{Method Overview}

Motivated by the above limitations, this paper introduces the Physics-Informed Harmonic Adaptive Spectral Transform Network, PHAST-Net, a unified framework for ideal T--F representation estimation, together with the harmonic PHAST-Net and Spline-PHAST-Net variants for harmonic suppression, noise reduction, component rerendering and reconstruction, and refined idealised Tempogram and Metrogram representations. PHAST-Net learns a direct mapping from a curated constellation of input observed Continuous Log-frequency Adaptive Wavelet Transforms (CLAWTs) to a high-fidelity, noise-suppressed, and cross-term-free idealised target representation via a high-capacity attention-guided network. The CLAWT constellation is selected through WVD-kernel analysis to provide representative coverage of local orientation and scale over the proposed T--F grid. Unlike prior learned T--F representation estimators that predict representations without explicit transform-family consistency, PHAST-Net regularises ITFR prediction with a physics-informed CLAWT reprojection loss, promoting transform-consistent, cross-term-free reconstruction while regularising optimisation. 

Additionally, in contrast to prior work, we formulate the representation on a \textit{log-frequency} T--F grid encompassing a frequency range specifically tailored to speech, music, and wider audio applications, handled by a Spatial Feature Transform (SFT). Establishing the framework in the log-frequency domain additionally addresses a key limitation of conventional linear-frequency-based approaches for harmonic signals, since harmonic overtone structure is translation-consistent (equivariant) along the log-frequency axis. This formulation therefore enables the construction of the harmonic PHAST-Net which, in addition to cross-term suppression and auto-term sharpening, learns controllable suppression of harmonic structure whilst retaining fundamental trajectories, yielding a proposed Harmonic ITFR that is especially suited to speech and music. In particular, the harmonic PHAST-Net can be employed to estimate the idealised overtone- and cross-term-free Fundamental Tempogram, which in turn is required for the construction of the Metrogram. 

To account for the wide range of nonstationary signal structures encountered in practice, training is performed on a procedurally generated and effectively unbounded synthetic dataset of observed-constellation–target pairings, comprising randomly generated spline-defined component trajectories with time-varying amplitudes, simulated overtone structures, component intersections, and variable noise floors. This training formulation additionally enables strong robustness to noisy real-world extracts, with the input scaling providing practical control over the degree of denoising and harmonic suppression at inference time. 

Finally, we additionally introduce the Spline-PHAST-Net variant, which detects ridge structure via Frangi ridge detection, tracks components through time using a data association algorithm, refines sub-pixel positions and magnitudes through spline-fitted trajectory models, and enables per-component signal reconstruction estimates and re-rendering on arbitrary target T--F grids of interest. 

\vspace{-2mm} \section{Preliminaries}

In this section we introduce the relevant material for Section~\ref{Methodology}, including the relationship between wavelet transforms, Cohen's class distributions, and the Wigner-Ville Distribution (WVD), in addition to the Idealised T--F representation (ITFR) model employed for the PHAST-Net model.

\vspace{-2mm} \subsection{Wavelet Transform Properties}

We first establish the fixed-window case of the kernel identity used for CLAWT. For a fixed complex-valued window $\Omega(t)$ \cite{wavelet,cozens2025rift}, define the magnitude-squared transform $\Phi(\omega,t)$ by:
\begin{smallalign}
    \Phi{\left(\omega, t\right)} &= \left |\left[z\left(\tau\right) * \Omega^*{\left(\tau\right)} e^{j \omega \tau}\right](t) \right |^2 = \left |\left[z * W_{\omega}\right]{\left(t\right)} \right |^2,
\end{smallalign}
\noindent where $W_{\omega}{\left(t\right)}$ is the modulated analysis function for angular frequency $\omega$, and $*$ denotes the continuous-time convolution. A significant relationship that can be derived is that for the fixed-window case, $\Phi{\left(\omega, t\right)}$ can be expressed as a 2D convolution \cite{cozens2025rift, Nuttall1988}:
\begin{smallalign}
\label{WVD_CWT_REL}
    \Phi{\left(\omega, t\right)}
    &=  2 \pi \int_{\mathbb{R}^2} WVD_z\left(\omega_1, T\right) WVD_{\Omega}\left(\omega - \omega_1, t - T\right) \, d T \, d \omega_1 \notag \\
    &= 2 \pi \left[WVD_z * \Pi \right] \left(\omega, t\right),
\end{smallalign}
\noindent where $\Pi(\omega, t)$ is the convolution kernel defined by $\Pi(\omega, t) \triangleq WVD_{\Omega}(\omega, t)$, and $WVD_z\left(\omega, \, t\right)$ and $WVD_{\Omega}\left(\omega, \, t\right)$ denote the Wigner--Ville Distributions (WVDs) of the signal, $z(t)$, and the analysis window, $\Omega{\left(t\right)}$, respectively. The WVD of the signal $z(t)$ is defined as \cite{wvd, ville}:
\begin{smallalign}
    WVD_z\left(\omega, \, t\right) = \frac{1}{2\pi}\int_{\mathbb{R}} z\left(t + \frac{\tau}{2}\right) z^*\left(t - \frac{\tau}{2}\right) e^{-j \omega \tau} \, d\tau.
\end{smallalign}
\noindent The analytic signal, $z(t)$, is computed as {\small $z(t) = x(t) + j\mathcal{H}\{x(t)\}$}, where $x(t)\in \mathbb{R}$ is the input signal, and $\mathcal{H}\{x(t)\}$ is the Hilbert transform \cite{ref:hilbert} of $x(t)$. Significantly, this relationship demonstrates that the fixed-window transform, $\Phi{\left(\omega, t\right)}$, is a Cohen's class bilinear distribution with a Cohen's class kernel $\Pi(\omega, t)$ \cite{Cohen1989, cozens2025rift}. For further details and derivations, please consult \cite{cozens2025rift}. Note that for the frequency-dependent CLAWT introduced in Section~\ref{wavelet_constellation}, the corresponding WVD kernel depends on the output frequency, providing a spatially varying extension of this fixed-window identity.

\vspace{-2mm} \subsection{Ideal Time-Frequency Representation (ITFR) Model}
\label{sec:itfr}

The concept of an Ideal T--F Representation (ITFR) for a signal is explored in several papers \cite{Mohammad, Boashash2016, cozens2025rift}. Let $z\left(t\right)$ be a multi-component complex signal \cite{Mohammad}:
\begin{smallalign}
\label{signal}
    z\left(t\right) &= \sum_{p=1}^P \mathds{1}_{{\cal \tau}_p}\!\left(t\right)A_p(t)\exp{\left(j\phi_{p, 1} + j \textstyle\int_{t_{p, 1}}^{t} \omega_p\left(\tau\right) \, d \tau\right)}
\end{smallalign}
\noindent where $A_p(t)$ are the positive real amplitudes for component {\small $p$ ($1 \le p \le P$)} respectively, and $\phi_p\left(t\right)$ are the phases of each component relative to the initial phase offsets, such that {\small $\phi_p(t) = \textstyle \int_{t_{p,1}}^{t} \omega_p(\tau)\, d\tau$}, where $\omega_p(t)$ are the instantaneous angular frequencies for component $p$. Here, $\phi_{p,1}$ denotes the initial phase at the onset $t_{p,1}$ of component $p$. The term $\mathds{1}_{{\cal \tau}_p}$ is an indicator function that confines the component to its active interval between its onset $t_{p, 1}$ and offset $t_{p, 2}$, with ${\cal \tau}_p=[t_{p,1},t_{p,2}]$. Conceptually, the ITFR consists of $P$ trajectories in the T-F plane with perfect time-frequency resolution, and no cross terms \cite{Mohammad, Boashash2016}. A suitable definition for the ITFR of a signal $z\left(t\right)$ in the \textit{linear frequency} domain (with respect to $\omega$), $\mathrm{ITFR}_{\omega}$, is thus:
\begin{smallalign}
\label{ITFR_linear}
    \mathrm{ITFR}_{\omega}\{z\}\left(\omega, \, t\right) = \sum_{p=1}^P \mathds{1}_{{\cal \tau}_p}\!\left(t\right)A_p(t)\delta\left(\omega - \omega_p(t)\right).
\end{smallalign}
\noindent Now let the angular frequency of the $s$th musical semitone be:
\begin{smallalign}
\label{semi}
    \omega_{s}(s) &= 2 \pi f_{0} \cdot 2^{\frac{s}{12}},
\end{smallalign}
\noindent where $f_0$ corresponds to the frequency of note $A0$ ($f_0 = 27.5 \si{Hz}$). Taking the coordinate system transformation $\omega \mapsto s$:
\begin{smallalign}
\label{ITFR_Eq_linear}
    \mathrm{ITFR}_{\omega}\{z\} \! \left(\omega_s(s), t\right)\! &= \!\left | \tfrac{\partial \omega_{s}(s)}{\partial s}\right |^{-1}\! \sum_{p=1}^P \mathds{1}_{{\cal \tau}_p}\!\left(t\right)A_p(t)\delta\left(s-s_p(t)\right) \notag\\[-0.5em]
    &=k_{f_0} \,2^{-\tfrac{s}{12}} \cdot\mathrm{ITFR}_s\{z\}(s,t)
\end{smallalign}
\noindent where {\small$k_{f_0}=\left(\tfrac{1}{12} 2 \pi f_0 \ln{2}\right)^{-1}$} is the scaling constant, {\small$s_p(t)=\omega_{s}^{-1}(\omega_p(t))$} (where $\omega_{s}^{-1}(\cdot)$ denotes the inverse of Eq.~\eqref{semi}), and $\mathrm{ITFR}_s$ is the \textit{logarithmic (semi-tonal)} $s$ domain ITFR:
\begin{smallalign}
\label{ITFR_log}
    \mathrm{ITFR}_s\{z\}(s,t)&=\sum_{p=1}^P \mathds{1}_{{\cal \tau}_p}\!\left(t\right)A_p(t)\delta\left(s-s_p(t)\right).
\end{smallalign}
\vspace{-4mm} \subsection{Harmonic ITFR Model}
\label{Harmonic_ITFR_Model}

Let $z_{H}\left(t\right)$ be a multi-component harmonic complex signal with a T--F varying (inhomogeneous) linearly distributed harmonic distribution:
\begin{smallalign}
\label{signal_harmonic_ITFR}
    &z_H\!\left(t\right) \!= 
    \!\!\sum_{p=1}^P \!\sum_{n_h=1}^{N_{H,p}} \!\!\mathds{1}_{{\cal \tau}_p}\!\left(t\right)\!A_h\!(n_h, p, t) A_p\!(t)\exp\!{\left(j \!\left[\phi_{p,n_h, 1} \!+ \!n_h \!\cdot \!\phi_{p}\!\left(t\right)\right]\right)}.
\end{smallalign}
\noindent where $A_h(n_h, p, t)$ is the relative amplitude of the $n_h$th harmonic, with $1 \le n_h \le N_{H,p}$; $N_{H,p}$ is the total number of harmonics for component $p$. Here, $\phi_{p,n_h,1}$ denotes the initial phase at the onset $t_{p,1}$ of component $p$ and harmonic $n_h$, such that as before $\phi_{p}(t) = \int_{t_{p,1}}^{t} \omega_p(\tau)\, d\tau$. Note that Eq.~\eqref{signal_harmonic_ITFR} is defined such that:
\begin{smallalign}
\label{norm_harmonics}
    \sum_{n_h=1}^{N_{H,p}}A_h(n_h, p, t)=1 \quad \quad\forall p, t
\end{smallalign}
\noindent Conceptually, the Harmonic ITFR likewise consists of $P$ trajectories in the T-F plane with perfect time-frequency resolution, and no cross terms. Thus, a suitable definition for the \textit{linear} $\omega$ domain Harmonic ITFR, for signal $z_{H}\left(t\right)$ is:
\begin{smallalign}
\label{Harmonic_ITFR_linear}
\mathrm{ITFR}_{H, \omega}\{z_H\}\left(\omega, \, t\right) &= \sum_{p=1}^P \mathds{1}_{{\cal \tau}_p}\!\left(t\right)A_p(t)\delta\left(\omega - \omega_p(t)\right).
\end{smallalign}
\noindent Note that the harmonic amplitudes, $A_h(n_h, p, t)$, disappear owing to the normalisation condition provided in  Eq.~\eqref{norm_harmonics}. By applying the same coordinate transformation $\omega \mapsto s$ introduced in Eq.~\eqref{ITFR_Eq_linear}, the corresponding Harmonic ITFR in the \textit{logarithmic (semi-tonal)} domain, $\mathrm{ITFR}_{H,s}\{z_{H}\}$, can be formulated as follows:
\begin{smallalign}
\label{Harmonic_ITFR_log}
    \mathrm{ITFR}_{H,s}\{z_{H}\}(s, t) &= \sum_{p=1}^P \mathds{1}_{{\cal \tau}_p}\!\left(t\right)A_p(t)\delta\left(s-s_p(t)\right)
\end{smallalign}
\noindent Note also that the logarithmic $s$ domain ITFR, $\mathrm{ITFR}_{s}$ (Eq.~\eqref{ITFR_log}), for signal $z_{H}\left(t\right)$ would likewise be:
\begin{smallalign}
\label{ITFR__h_Eq2}
    &\mathrm{ITFR}_{s}\{z_{H}\}\!\left(s, t\right) \notag \\
    &\quad= \!\sum_{p=1}^P \!\sum_{n_h=1}^{N_{H,p}} \!\!\mathds{1}_{{\cal \tau}_p}\!\left(t\right) \!A_h(n_h, p, t) A_p(t) \delta\left(s \!- \!s_p(t) \!- \!s_h(n_h)\right),
\end{smallalign}
\noindent where $s_h(n_h)=12 \log_2\!{n_h}$. As shown in the Supplementary Material (SM), Section~\ref{Harmonic_ITFR_properties}, redefining the Harmonic ITFR in the logarithmic domain renders the relationship between the ITFR and the Harmonic ITFR convolutional, thereby motivating its use in the proposed Harmonic ITFR estimation method. The analytical ITFR definitions above are expressed in the linear-amplitude domain; for network training and physics reprojection, their discrete rasterisations are constructed in the corresponding quadratic domain, as described in Section~\ref{data}.

\vspace{-2mm} \section{Proposed Methodology}
\label{Methodology}

\vspace{-1mm} \subsection{The Continuous Log-frequency Adaptive Wavelet Transform (CLAWT)}
\label{wavelet_constellation}

In this section, we specify the proposed constellation of wavelet transforms employed in the PHAST-Net framework, together with the corresponding spatially varying WVD kernels. Given that the framework is specified in log-frequency T-F coordinates, a wavelet with an exponentially varying instantaneous frequency (IF) is desired so that its IF trajectory is linear in this coordinate system. However, the principal-axis orientation $\theta$ and spread $\sigma$ of the induced kernel do not in general match the wavelet's IF direction and envelope width, as observed later in Eq.~\eqref{approx_Pi}. To obtain direct control of the desired $\theta$ and $\sigma$, we introduce the \emph{Continuous Log-frequency Adaptive Wavelet Transform} (CLAWT), a parametrisable wavelet transform family designed to provide controlled orientation and scale coverage in the log-frequency T--F plane.

The (magnitude-squared) CLAWT, $\Phi$, is defined in terms of a \textit{complex} wavelet function, $\Psi$, as follows: 
\begin{smallalign}
\label{ex:Phi}
   \Phi_{\sigma,\theta}{\left(s, t\right)} &= \left | \left[z * \Psi_{\sigma, \, \theta}^{(s)}\right]{\left(t\right)} \right |^2 \notag \\
   &= \left |\int_{\mathbb{R}} z\left(\tau\right) \,\Psi_{\sigma, \, \theta}^{(s)}{\left(t-\tau\right)} \, d \tau \right |^2\,.
\end{smallalign}
\noindent Here, $\Psi_{\sigma,\theta}^{(s)}(t)$ denotes the wavelet associated with semitone $s$, parameterised by the desired kernel principal-axis orientation $\theta$, measured relative to the time axis in this T--F coordinate system, and the relative spread parameter $\sigma$.

The corresponding CLAWT wavelet function is given by:
\begin{smallalign}
&\Psi_{\sigma, \, \theta}^{(s)}\left(t\right) \triangleq \psi_{\sigma_{in}(\sigma, \theta), \, \theta_{in}(\sigma, \theta)}^{(s)}(t) \label{wavelet}\\
\text{with:}\,\,&\psi_{\sigma, \, \theta}^{(s)}(t) = \frac{1}{\sqrt[4]{\pi \sigma_{s}^2}} e^{-\tfrac{t^2}{2\sigma_{s}^2}}\exp{\left[j\frac{\omega_{s}(s)}{\gamma} \left(e^{\gamma t}-1\right)\right]}, \notag\\ &\gamma \!=\! \frac{\ln{2}}{12}\tan{\theta},\,\,  \,  
    \sigma_s \!=\! \frac{\sigma}{\sqrt{k_s}}, \,\,\,   k_s \!=\!2\pi f_0\,2^{s/12}\left(\frac{\ln2}{12}\right)\!.\label{component_wavelet}
\end{smallalign}
\noindent Here, $\sigma_{in} \equiv \sigma_{I} \, \sigma_0(\sigma,\theta)$ and $\theta_{in} \equiv -\arctan\left(\tan{\left(\kappa(\sigma,\theta)\right)}/\sigma_{I}^2\right)$. The reference scale $\sigma_I$ is chosen such that the Gaussian approximation of the kernel is isotropic for $\sigma=1$ under the sampled T--F grid. At $\gamma = 0$, the appropriate limit defines the phase as $\omega_s(s) t$. To match the desired principal-axis scale and orientation of this local approximating Gaussian kernel, the wavelet parameters are obtained using a covariance-matching transformation $(\sigma, \theta) \mapsto (\sigma_0, \kappa) $ as in \cite{cozens2025rift}:
\begin{smallalign}
        \sigma_{0}(\sigma,\theta)
  \!&= \sqrt{\sigma^2\cos^2\theta+\sigma^{-2}\sin^2\theta} \label{sigma_{kappa}}\\
    \kappa(\sigma,\theta)&=\operatorname{atan2}\!\left((\sigma^2\!-\!\sigma^{-2})\sin\theta\cos\theta,\sigma_{0}^2(\sigma,\theta)\right).\label{phi_core}
\end{smallalign}
\noindent Due to the wavelet's rotational symmetry, $\Psi_{\sigma,\theta}^{(s)}(t)=\Psi_{\sigma,\theta \pm \pi}^{(s)}(t)$, for $|\theta| > \tfrac{\pi}{2}$. Note also that by design the instantaneous semitonal frequency of $\psi_{\sigma,\theta}^{(s)}$, $s_{\text{inst}}$, is:
\begin{smallalign}
    s_{\text{inst}}\!&= \!\omega_{s}^{-1}\!(\omega_{\text{inst}})
    \!=\! \omega_{s}^{-1}\!\left(\!2 \pi f_0 2^{{\tfrac{1}{12}\tan{\theta} t} + \tfrac{s}{12}}\right) \!=\! s \!+\! \tan{(\theta)} t,
\end{smallalign}
\noindent where $\omega_{s}^{-1}$ is the inverse of Eq.~\eqref{semi}. Thus, the IF trajectory of the base wavelet function $\psi_{\sigma, \, \theta}^{(s)}\left(t\right)$ is a straight line with slope $\tan{\theta}$ in the $(s, t)$ plane. To derive the corresponding CLAWT kernel in the log-frequency coordinate system, Eq.~\eqref{ex:Phi} can be expressed using a spatially varying extension of the identity in Eq.~\eqref{WVD_CWT_REL}:
\begin{smallalign}
   \Phi_{\sigma,\theta}{\left(s, t\right)} &= 2\pi\int_{\mathbb{R}^2} WVD_z\left(\omega_{s}(\upsilon), \tau\right) \cdot \Pi_{\sigma, \theta}^{(s)}(s - \upsilon,t - \tau)\, d \upsilon \, d \tau
\end{smallalign}
\noindent Thus, taking into account the coordinate system transformation $\omega \mapsto s$, the following \emph{spatially-variant} kernel can be derived:
\begin{smallalign}
&\Pi_{\sigma, \theta}^{(s)}\left(\upsilon, \tau\right) \triangleq
\bar{\Pi}_{\sigma_{in}(\sigma, \theta), \theta_{in}(\sigma, \theta)}^{(s)}\!\left(\upsilon, \tau\right)\label{pi_relationship}\\
   &\text{with:}\,\,\bar{\Pi}_{\sigma, \theta}^{(s)}(\upsilon,\,\tau) = f_0 \ln{2}\cdot 2^{\frac{s-\upsilon}{12}}\frac{e^{-\tau^{2}/\sigma_{s}^{2}}}{12\sqrt{\pi}\sigma_s}\notag\\
   &\cdot\! 
    \int_{\mathbb{R}}
      \!\exp\!\left[-\frac{\xi^{2}}{4\sigma_{s}^{2}}
          \!-\!j\omega_{s}(s\! -\! \upsilon) \xi + j\frac{\omega_{s}(s)}{\gamma}e^{\gamma \tau}\,
        2\sinh\!\left(\tfrac{\gamma\xi}{2}\right)\right]\,d\xi.\label{exact_kernel_Pi}
  \end{smallalign}
\noindent The full derivation is provided in Section~\ref{derivation_kernel} of the SM. The calibrated exact kernel $\Pi_{\sigma, \theta}^{(s)}$ (Eq.~\eqref{pi_relationship}) is used for the physics reprojection in Section~\ref{framework}. The local Gaussian approximation to the base kernel, $\hat{\Pi}_{\sigma, \theta}^{(s)} \approx \bar{\Pi}_{\sigma, \theta}^{(s)}$, is obtained from Eq. \eqref{exact_kernel_Pi} as follows:
\begin{smallalign}
\label{approx_Pi}
&\hat{\Pi}_{\sigma, \theta}^{(s)}(\upsilon,\tau) \!\propto \!\frac{f_0 \ln{2}}{6}\!\cdot\!\exp\!\left[\!-k_{s}\!\left(\frac{\tau^{2}}{\sigma^{2}}
         \!+\!\sigma^{2}\left(\upsilon \!+ \!\tau \cdot \tan{\theta}\right)^{2}\right)
   \right]\!.
\end{smallalign}
\noindent Thus, this approximation is a skewed Gaussian, whose standard deviation along any axis decreases with increasing frequency proportionally to $k_s^{-1/2}$, motivating the calibration in Eq.~\eqref{wavelet} to ensure that the parameters $\sigma$ and $\theta$ specify the desired kernel principal-axis scale and orientation, rather than directly controlling the wavelet envelope width and IF slope. Note that for a discretised T--F coordinate system with step size $(\Delta s, \Delta t)$, the unchirped base kernel in Eq.~\eqref{approx_Pi} is isotropic at the reference scale $\sigma =\sigma_I=\sqrt{\Delta t / \Delta s}$, corresponding to $\sigma=1$ under the calibrated CLAWT parametrisation.

To select a suitable constellation, an appropriate region of $\sigma$ and $\theta$ must be represented to promote auto-term isolation and cross-term suppression across the constellation for a variety of signal types with varying chirp angles and curvatures. In a previous publication \cite{cozens2025rift}, a uniformly distributed $\theta$ and log normally distributed $\sigma$ centred on the isotropic standard deviation $\sigma_I$ were proposed. Here, to balance computational efficiency and coverage, we propose a $K=15$ constellation comprising one isotropic member and 14 anisotropic members with uniformly spaced orientations, ordered as $\sigma_k=[1,7,\dots,7]$ and $\theta_k=[0,f_\theta(1), f_\theta(8),f_\theta(2),\dots,f_\theta(7), f_\theta(9),\dots,f_\theta(14)]$, where:
\begin{smallalign}
f_\theta(k)=\frac{\pi\left(k-8\right)}{14}.
\end{smallalign}
\noindent Note that this specific ordering groups the isotropic and axis-aligned members (1--3) before the obliquely oriented members (4--15). Panels (a)--(e) and (g)--(k) of Fig.~\ref{fig:evaluation_constellation} in the SM show selected members of an example input CLAWT constellation corresponding to the waveform generated from the ITFR in panel (f).

\vspace{-0mm} \subsection{PHAST-Net framework}
\label{framework}

\begin{figure*}
    \centering
    \includegraphics[width=1.0\textwidth,
    trim={0.0cm 0.5cm 0.0cm 0.7cm}, %LBRT
        clip]{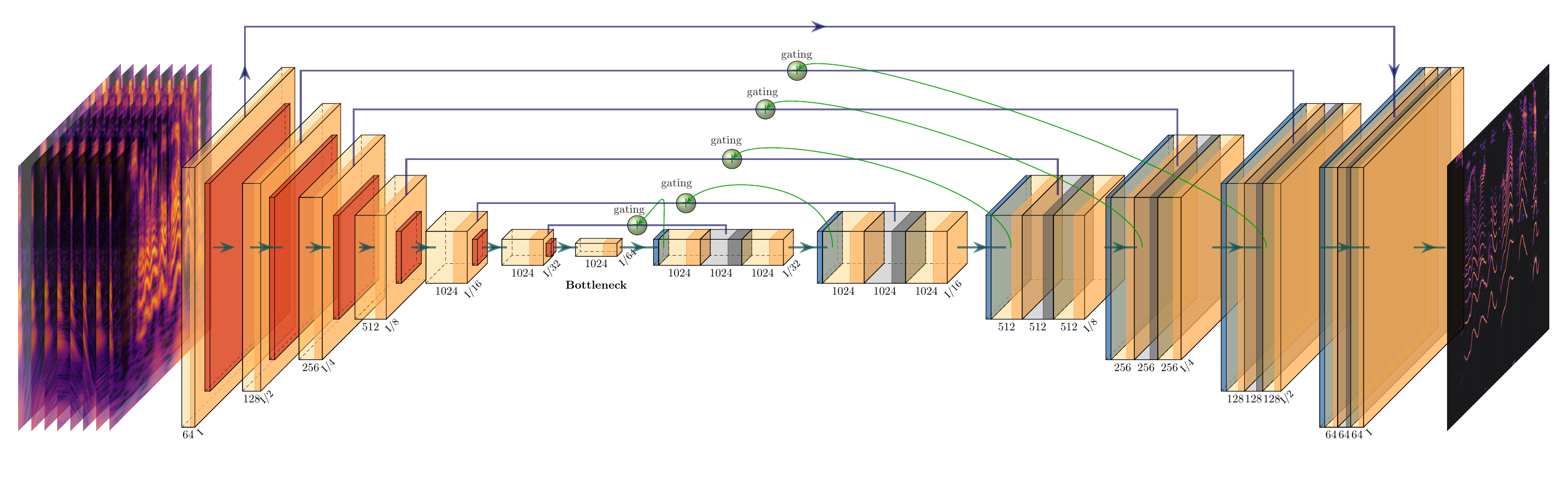}
    \vspace{-12mm}\caption{PHAST-Net Architecture \textit{(I = 1024 denotes the input spatial dimension)}.}
    \label{fig:rift_net_arch}\vspace{-4mm}
\end{figure*}

This section establishes the core components of the PHAST-Net framework, including the ITFR and Harmonic ITFR estimation process. As noted in the overview, while classical methods exist for refining T--F representations, the integration of a neural architecture is motivated by its ability to provide a single-pass deconvolution solution, as well as to perform automatic auto-term extraction \cite{Zhang2022CTFreeCNN}. In addition, typical classical methods employ the cross-term suppressed WVD as a target for the ITFR \cite{Cohen1989}, which is conceptually aligned but not equivalent to the ITFR model, due to the presence of artificial auto-terms, further supporting the employment of neural architectures to enable refined auto-term extraction. Motivated by the strong performance of the classical RIFT algorithm \cite{cozens2025rift} and recent deep learning approaches \cite{Pan2023TFANet, Zhang2022CTFreeCNN}, we develop a deep-learning framework for joint deconvolution and auto-term extraction. Driven by the success of attention-guided autoencoder architectures in adjacent fields \cite{UNET}, we propose a deep, high-capacity attention-guided U-Net-style architecture which takes as input a constellation tensor of Continuous Log-frequency Adaptive Wavelet Transforms (CLAWTs) and outputs the estimate of the proposed logarithmic-frequency domain ITFR and Harmonic ITFR (Eq.\eqref{ITFR_log} and Eq.\eqref{Harmonic_ITFR_log}). The architecture is presented in Fig.~\ref{fig:rift_net_arch}.

To account for the frequency-dependent CLAWT kernel induced by the logarithmic-frequency scaling, the network incorporates a frequency-component Spatial Feature Transform (SFT) \cite{SFT}, which provides spatially adaptive affine parameter modulation. Likewise, in contrast to recent learned T--F representation estimators, which predict representations without explicit consistency with the underlying transform family, PHAST-Net learns a direct transform-consistent mapping through the incorporation of a regularising physics-informed auxiliary loss. Specifically, during training, the predicted and target ITFRs are reprojected through the derived CLAWT kernels, and the auxiliary loss penalises discrepancies between these reprojections, thereby promoting transform and energy-distribution consistency and regularising optimisation through denser auxiliary targets.

Let $n=1,\ldots,N$ index samples within a training batch, and let the time-frequency image dimensions be $H \times W$, with pixel indices $(i, j)$. The ITFR (tonal) and Harmonic models are each trained on dedicated datasets, as detailed in Section~\ref{data}. Table~\ref{tab:phast_notation} summarises the principal tensors and operators employed in the PHAST-Net framework. 

\begin{table}[t]
\caption{Principal PHAST-Net tensors and operators.}
\label{tab:phast_notation}
\centering
\footnotesize
\setlength{\tabcolsep}{3pt}
\vspace{-2mm}\begin{tabular}{p{0.15\linewidth}p{0.80\linewidth}}
\hline
Symbol & Definition \\
\hline

$\mathbf{\Phi}_n$ &
Input CLAWT constellation,
$K\times H\times W$
(Section~\ref{wavelet_constellation}). \\

$\mathbf{x}_n,\hat{\mathbf{x}}_n$ &
Target and predicted quadratic-domain ITFR rasterisations,
$H\times W$, corresponding to tonal or harmonic models
(Section~\ref{data}). \\

$\mathbf{y}_n,\hat{\mathbf{y}}_n$ &
Target and predicted physics reprojections,
$\mathcal{P}(\mathbf{x}_n)$ and
$\mathcal{P}(\hat{\mathbf{x}}_n)$,
each $K\times H\times W$
(Eq.~\eqref{kernels}). \\

$f_{\mathbf{\Theta}}$ &
PHAST-Net, with trainable parameters $\mathbf{\Theta}$;
$\hat{\mathbf{x}}_n=f_{\mathbf{\Theta}}(\mathbf{\Phi}_n)$. \\

$\mathcal{P},\mathbf{\Pi}_k$ &
Spatially varying physics reprojection operator and its
$k$th discretised CLAWT kernel
(Eq.~\eqref{kernels} and~\eqref{pi_relationship}). \\

\hline
\end{tabular}\vspace{-3mm}
\end{table}

 The CLAWT constellation tensor, $\mathbf{\Phi}_n$, is constructed by stacking $K$ discretised input CLAWTs, each defined by a specific orientation--scale pair in the T--F plane, as described in Section~\ref{wavelet_constellation}. Note that at inference, prior to CLAWT computation, the observed waveform $z[m]$ is RMS-normalised and subsequently scaled by a parameter $\alpha_{\mathrm{noise}}$, which controls the effective level of noise suppression (as described in Section~\ref{data}), such that $z_{\mathrm{norm}}[m]=\alpha_{\mathrm{noise}}\cdot z[m]/RMS(z)$. \textbf{$\mathcal{P}$} is the physics operator, such that $\mathbf{\hat{y}}_{n,k} = \mathbf{\hat{x}}_n *_{SV} \mathbf{\Pi}_k$ for $\mathbf{\hat{y}}_n=\{\mathbf{\hat{y}}_{n, k}\}_{k=1}^K$, where $*_{SV}$ denotes 2D spatially varying discrete convolution over the T--F grid with the corresponding spatially varying $k$th CLAWT kernel, $\mathbf{\Pi}_k$, such that:
\begin{smallalign}
\label{kernels}
\hat{y}_{n,k;i,j}
= \left[\mathbf{\hat{x}}_n *_{SV} \mathbf{\Pi}_k\right]_{i,j}
&= \sum_{u,v}\hat{x}_{n;i-u,j-v}\,\Pi_{k;u,v;i,j},
\end{smallalign}
\noindent where $\mathbf{\hat{x}}_n=\{\hat{x}_{n;i, j}\}_{i=1,\dots,H, \,j=1,\dots,W}$ and $\mathbf{\Pi}_k=\{\Pi_{k; u, v; i, j}\}_{u=1,\dots,U_{\Pi}, \,v=1,\dots,V_{\Pi};i=1,\dots,H, \,j=1,\dots,W}$, with $(U_{\Pi},V_{\Pi})$ denoting the support of each of the kernels in $\mathbf{\Pi}_k$; the specific forms of the CLAWT kernels $\mathbf{\Pi}_k$ are described in Section~\ref{wavelet_constellation}. The proposed loss function is then given by:
\begin{smallalign}
\label{eq:loss}
\mathcal{L}_{\text{total}}&=\mathcal{L}_{x,\text{mse}} + \mathcal{L}_{x,\text{log}} + \mathcal{L}_{y,\text{mse}} + \mathcal{L}_{y,\text{log}}+\mathcal{L}_{TV}.
\end{smallalign}
\noindent Both domains use MSE and frequency-weighted log-MAE losses, with masks $\mathbf W_x$,$\mathbf W_y$ emphasising lower frequencies and $\epsilon$ controlling logarithmic sensitivity. Total variation penalises spatial differences in the predicted ITFR. Full definitions and normalisations are provided in SM Section~\ref{sec:loss_details}.

\begin{figure*}
    \centering
    \includegraphics[width=1.0\textwidth,
    trim={0.0cm 0.0cm 0.0cm 0.0cm}, %LBRT
        clip]{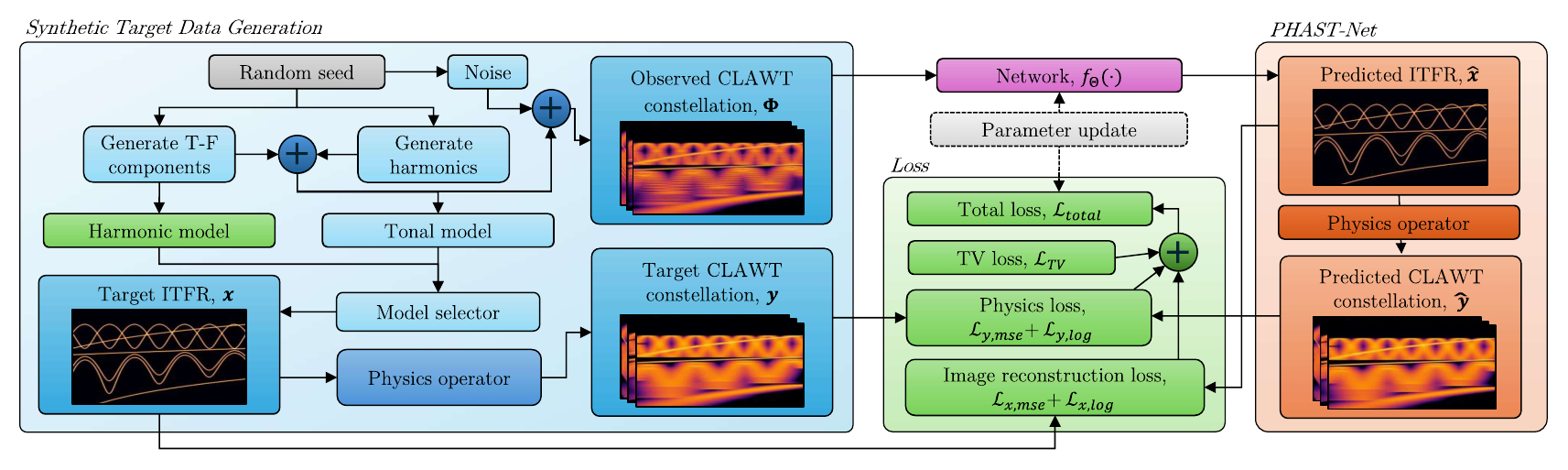}
    \vspace{-8mm}\caption{Block diagram representing the PHAST-Net training process}
    \label{fig:block_diagram_training}\vspace{-4mm}
\end{figure*}

Both the tonal and harmonic models were trained using the Adam optimiser \cite{Kingma2015Adam}, with a batch size of $16$ and an initial learning rate of $10^{-4}$, which was subsequently reduced during training until convergence. Approximately $40\,000$ procedurally generated training examples were used per model. Training was performed on a single NVIDIA RTX 4090 GPU, with cumulative training time of approximately one month. The harmonic model was initialised with the converged weights of the tonal model and trained on the corresponding harmonic input-target pairs for the same underlying examples. A schematic of the PHAST-Net training process is shown in Fig.~\ref{fig:block_diagram_training}.

\vspace{-2mm} \subsection{Generation of Synthetic Training Data}
\label{data}

\begin{figure*}
    \centering
    \vspace{-0mm}\includegraphics[
        width=1.0\textwidth,
        trim={0.25cm 0.45cm 0.3cm 0.7cm}, %LBRT
        clip
    ]{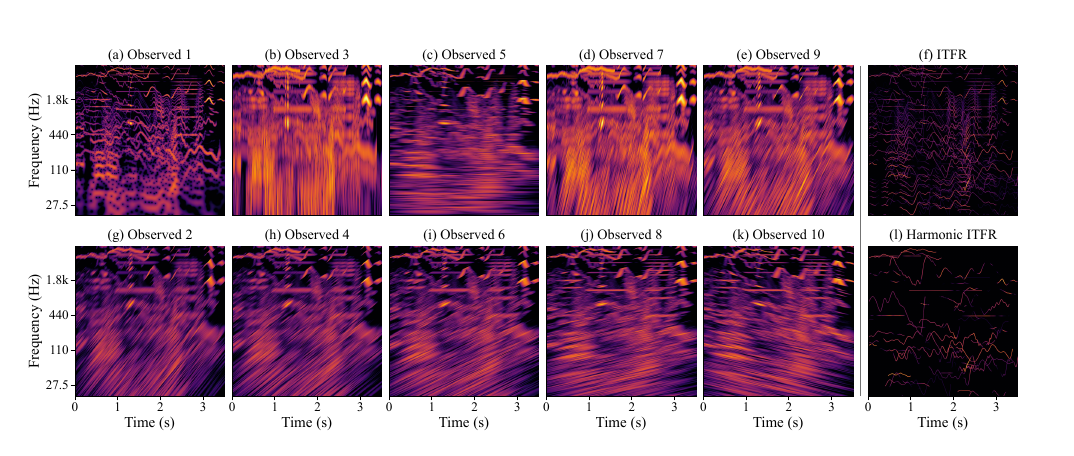}
    \vspace{-3mm}
    \caption{Example procedurally generated input data point, comprising a randomly generated tonal ITFR target in Panel (f), the corresponding harmonic ITFR target in Panel (l), and selected members of the input CLAWT constellation (members 1-10). The heatmap depth is $-90\,\mathrm{dB}$.}
    \label{fig:evaluation_constellation}\vspace{-0mm}
\end{figure*}

 Since the neural inference task corresponds to learning an underlying physical process, the required training data can be generated procedurally on demand. Consequently, the use of synthetic training data makes the effective dataset size practically unbounded. The $n$th training data sample, $\mathbf{z}_n=\{z_{n,m}\}_{m=1,\dots,M}$ (where $m$ denotes the time index) is procedurally generated as the sum of randomly generated sinusoidal components, with instantaneous frequencies and magnitudes specified by randomly generated spline functions \cite{deBoor1978}. The ITFR target, $\mathbf{x}_n$, is generated by rasterising the idealised trajectories onto the $H \times W$ T--F grid. Although the analytical ITFR is defined in the linear-amplitude domain, the component amplitudes are squared prior to Xiaolin Wu rasterisation, ensuring that the resulting discrete training targets reside in the same quadratic domain as the magnitude-squared CLAWT observations. To rasterise the T--F components, the antialiased line algorithm \cite{wu1991aa} is employed. The corresponding input constellation, $\mathbf{\Phi}_n$ is thus derived from taking the $K$ CLAWTs, parametrised by $\{\theta_k\}_{k=1}^K$ and $\{\sigma_k\}_{k=1}^K$ as outlined in Section~\ref{wavelet_constellation}. To improve robustness to varying noise levels, Gaussian noise is added to $\mathbf{z}_n$, where the standard deviation is randomly drawn from a prescribed interval independently for each data point in the dataset. Consequently, scaling the input constellation during inference enables noise suppression to be dynamically controlled, while simultaneously providing control over the degree of harmonic suppression in the harmonic model. Additionally, for each component, a randomly determined number of synthesised harmonic partials is included, with varying amplitudes and linearly spaced overtone positions, following the formulation given in Section~\ref{Harmonic_ITFR_Model}. Small amounts of jitter are introduced to enhance robustness to deviations from strictly linear harmonic spacing. Thus, for the $n$th data point in the dataset, a harmonic-suppressed target for the Harmonic PHAST-Net can be generated by only rasterising the fundamentals onto the target $\mathbf{x}_n$ T--F grid, with the linear-domain component-amplitude parameters $A_p(t)$ squared before rasterisation in the same manner as the tonal targets. To illustrate the process, Panels (f) and (l) of Fig.~\ref{fig:evaluation_constellation} in the SM show procedurally generated tonal and harmonic ITFR targets, respectively.

\vspace{-2mm} \subsection{Fundamental Tempogram ITFR Estimation}
\label{sec:fundamental_tempogram}

As introduced in the overview, a Tempogram is commonly obtained by applying T--F analysis to an audio-derived onset-detection (novelty) function \cite{ref:tempogram1,ref:tempogram2, Gui2018TempogramMP, Cozens}. In many applications, harmonic suppression is highly desirable; we use the term \emph{Fundamental Tempogram}, introduced in  \cite{Cozens}, to denote this idealised, fundamental-only representation, which can be expressed in terms of the Harmonic ITFR (Eq.~\eqref{Harmonic_ITFR_log}):
\begin{smallalign}
\label{fundamental_tempogram}
    \text{T}_{F, \lambda}\left\{z_H\right\}(\lambda, \, t) &= \mathrm{ITFR}_{H,s}\left\{f_{\text{onset}}\left\{z_H\right\}(t)\right\}(\lambda, t),
\end{smallalign}
\noindent where $\lambda$ is the logarithmic semitonal tempo ($\lambda = s$), and the proposed idealised onset-detection (novelty) function is:
\begin{smallalign}
\label{z_onset}
    f_{\text{onset}}\left\{z_H\right\}(t) &= \left(\frac{\partial}{\partial t}\int_{\mathbb{R}}\mathrm{ITFR}_{s}\{z_H\}(s,t)\,ds\right)_+,
\end{smallalign}
\noindent with $(x)_+ = \max\left\{0, \, x\right\}$. In practice, this operation is applied directly to the predicted quadratic-domain ITFR. For the analytic amplitude-domain ITFR, $\int_{\mathbb{R}}\mathrm{ITFR}_{s}\{z_H\}(s,t)\,ds$ is analogous to the cross-term-free analytic envelope of the signal $z_H(t)$:
\begin{smallalign}
    &\int_{\mathbb{R}}\mathrm{ITFR}_{s}\{z_H\}(s,t)\,ds \notag\\&=\int_{\mathbb{R}}\left(\!\sum_{p=1}^P \!\sum_{n_h=1}^{N_{H,p}} \!\!\mathds{1}_{{\cal \tau}_p}\!\left(t\right) \!A_h(n_h, p, t) A_p(t) \delta\left(s \!- \!s_p(t) \!- \!s_h(n_h)\right)\right)\,ds\notag \\
    &= \!\sum_{p=1}^P \!\sum_{n_h=1}^{N_{H,p}} \!\!\mathds{1}_{{\cal \tau}_p}\!\left(t\right) \!A_h(n_h, p, t) A_p(t)= \!\sum_{p=1}^P\!\mathds{1}_{{\cal \tau}_p}\!\left(t\right) \!A_p(t),
\end{smallalign}
\noindent where the final step follows from Eq.~\eqref{norm_harmonics}. Thus, taking the gradient of this idealised analytic envelope extracts note onsets (transients) without cross-term interference, such as beating frequencies that would otherwise distort the representation. Likewise, taking positive components retains positive envelope changes associated with onsets and suppresses negative changes associated with note ends. 

\vspace{-3mm} \subsection{Metrogram ITFR Estimation}

The \emph{Metrogram} is a transform that encodes time-varying metric structure by applying the multiplicative analogue of the autocorrelation function to the Fundamental Tempogram in the linear frequency domain:
\begin{smallalign}
\label{ex:10}
    \mathcal{M}\left(k, \, t\right) &= \frac{1}{Z\left(k\right)}\int_{0^{+}}^{\infty} \text{T}_F\left(\omega_{bpm}, \, t\right) \text{T}_F\left(k\omega_{bpm}, \, t\right) \, d \omega_{bpm},
\end{smallalign}
\noindent where $\mathcal{M}\left(k, \, t\right)$ is the Metrogram, $k$ is the rhythmic ratio to be evaluated, $Z\left(k\right)$ is a normalising function, and $\text{T}_F\left(\omega_{bpm}, \, t\right)$ is the \emph{linear} tempo axis ($\omega_{bpm}$) Fundamental Tempogram ITFR. The Metrogram is constructed such that the factor $k$ provides insight into rhythmic ratios (metric information) present in the extract, independent of tempo, thereby enabling direct inference of time signature, polymetre, and polyrhythm \cite{Cozens}. For instance, for a Metrogram restricted to $k>1$, an idealised 4/4 extract will result in a peak at $k=4$ in the Metrogram ITFR, $\mathcal{M}\left(k, \, t\right)\propto\delta(k-4)$. Supplementary Material (SM), Section~\ref{metrogram_derivation} provides a derivation of this result.

For the logarithmic-tempo representation, we define the corresponding log-domain Metrogram directly as:
\begin{smallalign}
\label{metrogram_definition}
\mathcal{M}(k, t)
&= \int_{\mathbb{R}}\text{T}_{F,\lambda}(\lambda, t)\,
\text{T}_{F,\lambda}(\lambda + 12\log_2 k, t)\, d\lambda \\
&= R_{TT}\!\left(12\log_2 k,\, t\right),
\end{smallalign}
\noindent where $R_{TT}(x,t)$ denotes the autocorrelation of $\text{T}_{F,\lambda}(\cdot,t)$ along $\lambda$:
\begin{smallalign}
R_{TT}(x,t) = \int_{\mathbb{R}}\text{T}_{F,\lambda}(\lambda, t)\,
\text{T}_{F,\lambda}(\lambda + x, t)\, d\lambda.
\end{smallalign}
\noindent Note that $Z\left(k\right)$ is omitted here; it was introduced in \cite{Cozens} for probabilistic calibration across $k$, which is not required for the present analysis. 

\vspace{-2mm} \subsection{Spline-Based Component Reconstruction}
\label{sec:component_extraction}

\vspace{-1mm}This section describes the component-extraction stage that converts the anti-aliased ridge representation predicted by PHAST-Net into explicit spline-parameterised ridge components, which are subsequently phase-aligned to form the \textit{Spline-PHAST-Net} variant. Candidate ridge structures are first identified using the multiscale Frangi ridge detector \cite{Frangi1998Vessel}, after which thresholding and skeletonisation are applied to obtain a graph of dominant curvilinear features. Local amplitudes and sub-pixel centreline positions are recovered from the Xiaolin Wu line algorithm \cite{wu1991aa} interpolation weights, thereby retaining the sub-pixel information encoded by the PHAST-Net output. Ambiguous connectivity at junctions and branch points is then formulated as a data-association problem and resolved using the Hungarian algorithm \cite{kuhn1955hungarian}. Finally, the recovered ridge tracks are fitted with splines \cite{deBoor1978}, yielding smooth continuous curves that can be resampled, mapped, and rasterised to arbitrary T--F coordinates.

Waveforms are resynthesised from the recovered amplitudes and integrated instantaneous frequencies. A single phase offset per component is chosen to maximise correlation with the input waveform; the amplitude rescaling and closed-form solution are given in SM Section~\ref{sec:phase_recovery}.

\vspace{-2mm} \section{Results}

This section presents an empirical evaluation of the proposed PHAST-Net framework across a diverse set of tasks to demonstrate its effectiveness and versatility. In particular, we consider speech, music, and synthetic signal excerpts, as well as a challenging synthetic waveform benchmark, thereby enabling a comparative assessment of PHAST-Net against established methodologies. Given that exact ITFR ground truth is unavailable for natural recordings, controlled synthetic signals are used for reference-based quantitative evaluation, while speech and music recordings provide complementary evidence of real-world applicability.

We first analyse a speech signal taken from the Harvard sentences database~\cite{speech}, shown in Fig.~\ref{fig:evaluation_speech}. The figure depicts three selected members of the input CLAWT constellation (panels (a)–(c)) and the corresponding PHAST-Net–predicted cross-term-free CLAWT representations (panels (f)–(h)). The raw PHAST-Net outputs for the tonal and harmonic models are shown in panels (d) and (e), respectively, while the associated Spline-PHAST-Net outputs are provided in panels (i) and (j). Next, we examine a violin excerpt from the composition Piano Concerto No. 4 (by the author), illustrated in Fig.~\ref{fig:evaluation_music}, which exhibits pronounced frequency-modulated vibrato. The figure is structured analogously, showing selected members from both the input and predicted CLAWT constellations together with the corresponding PHAST-Net outputs. The specific notated score for the analysed excerpt is:

\vspace{-1mm}
\begin{figure}[H]
    \centering
    \vspace{-0mm}\includegraphics[
        width=1\columnwidth,
        trim={0.75cm 23.7cm 0.3cm 4.3cm}, %LBRT
        clip
    ]{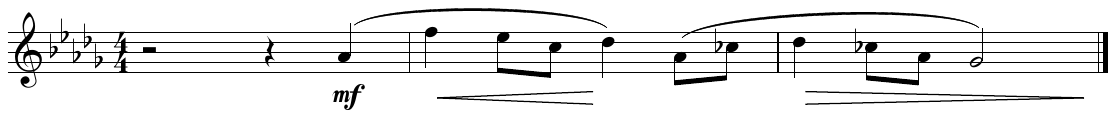}
    \vspace{-5mm}
    \label{fig:music}
\end{figure}\vspace{-1mm}

\noindent In addition, the PHAST-Net framework is evaluated on a complex, multi-component synthetic waveform characterised by substantial cross-term interference, as shown in Fig.~\ref{fig:evaluation_synthetic}. The extracted components (colour-coded) are presented in panel (j), and the waveform is defined by Eq.~\ref{eq:complex_signal_log} in the SM.

To further assess the PHAST-Net–predicted Fundamental Tempogram and Metrogram ITFR representations, we construct a controlled synthetic musical excerpt that incorporates both a time-varying tempo and a change of time signature. This design enables a direct comparison with idealised ground-truth Fundamental Tempogram and Metrogram representations. Specifically, the idealised excerpt consists of 32 measures in 3/4 metre followed by 32 measures in 4/4 metre, generated from a simple periodic chordal pattern. The tempo follows a linearly increasing trajectory, onto which a low-frequency sinusoidal modulation is superimposed to emulate pronounced rubato and to test the Metrogram’s robustness to substantial tempo fluctuations. As described in Section~\ref{sec:fundamental_tempogram}, the PHAST-Net Tempogram is computed from the note-onset detection function derived from the inferred PHAST-Net ITFR. The simulated idealised Fundamental Tempogram and normalised Metrogram ITFRs are given by:
\begin{smallalign}
&b(t)=187.5+\frac{187.5t}{T_2}+20\sin\!\left(\frac{\pi t}{5}\right),
\,
m(t)=3+\mathds{1}_{{\cal \tau}_m}(t),\notag\\
&T_{\!F}(\omega_{\rm bpm},t)
=\delta(\omega_{\rm bpm}-b(t))
+4\delta(\omega_{\rm bpm}-b(t)/m(t)),\label{ex:tempogram}\\
&\mathcal{M}(k,t)=\delta(k-m(t)),
\qquad
T_2=47.687{\rm s},\ T_1=24.239{\rm s},\label{ex:metrogram} 
\end{smallalign}
\noindent with ${\cal \tau}_m=[T_1,T_2]$. For figure visibility at publication scale, PHAST-Net outputs are rendered after applying a small $\sigma < 1$-pixel isotropic Gaussian anti-aliasing filter.

Subsequently, a comparative evaluation is conducted with respect to representative T--F analysis methods on a challenging synthetic multicomponent dataset specifically constructed to assess T--F localisation, interference suppression, and robustness to additive white Gaussian noise (AWGN). Specifically, the dataset consists of 15 independently generated spline-based test signals, each comprising multiple closely spaced fundamental components, rich simulated overtone structure, component intersections, and pronounced cross-term interference. Since the majority of established T--F refinement methods are formulated in the linear-frequency domain, the benchmark is evaluated on a common $1024 \times 1024$ linear-frequency T--F grid spanning $100-1000\,\mathrm{Hz}$. Given that the native PHAST-Net output is defined on a log-frequency grid,  quantitative comparisons use Spline-PHAST-Net rerendered directly onto the common linear-frequency grid, derived from the corresponding PHAST-Net output. The raw PHAST-Net output is also shown for reference, but not used for the linear-grid quantitative comparisons. 

We compare against a representative set of reproducible T--F analysis methods: the Synchrosqueezing transform (SST) \cite{Thakur2013}, classical reassignment (RS) \cite{reassignment}, the adaptive optimal-kernel (AOK) distribution \cite{jones2}, the S-Method \cite{stankovic1994smethod}, the Choi--Williams distribution (CW)
\cite{ChoiWilliamsDistribution}, the RIFT algorithm and its corresponding spline variant \cite{cozens2025rift}, and the synchroextracting transform (SET) \cite{yu2017set}. The comparison set was selected to include representative methods that can be evaluated reproducibly on the common sampled signals and prescribed T--F grid, excluding learned approaches for which direct evaluation would require retraining or architectural modification, or whose native frequency support and resolution are incompatible with the benchmark. Note that SST, SET, S-Method, and reassignment are computed from the same Gaussian STFT, whose smoothing kernel is isotropic relative to the normalised T--F grid. Likewise, all methods are evaluated on the same underlying noisy waveforms and common output T--F grid, and method-specific settings are fixed across test signals and SNR conditions (SM Table~\ref{tab:baseline_policy}). The input waveform for PHAST-Net is deterministically resampled to its required native sampling rate before CLAWT analysis.

The comparative benchmark employs three complementary metrics: (1) Bhattacharyya overlap (BC; higher is better) \cite{Bhattacharyya1943}, (2) Jensen--Shannon divergence (JS; lower is better) \cite{Lin1991}, and (3) ridge energy ratio (RER; higher is better) \cite{cozens2025rift}. The metrics compare each method with the ground-truth target ITFR, applying a soft Gaussian spatial tolerance ($\sigma_{\mathrm{eval}}=2$ pixels) to reduce sensitivity to sub-pixel discretisation. BC and JS provide measures of distributional fidelity, whilst the RER evaluates the proportion of energy within a soft neighbourhood of the target ITFR ridges, subject to the same spatial tolerance. The precise definitions, normalisations, and tolerance studies are provided in the SM, Section~\ref{sec:evaluation_metrics} and Table~\ref{tab:sigma_ablation}. 

The methods are evaluated at AWGN levels of $-5, 0$, and $5 \mathrm{dB}$, together with the noise-free case denoted by $\infty \mathrm{dB}$. Qualitative results at $0 \mathrm{dB}$ are shown in Fig.~\ref{fig:evaluation}, and results for all SNRs are reported in Table~\ref{tab:metrics_combined}. For conciseness, raw RIFT and the WVD are omitted from the main table, with Spline-RIFT retained as the better-performing RIFT variant; their comparative rankings are nevertheless reported in the SM ablation (Tab.~\ref{tab:sigma_ablation}). At each SNR, entries report the mean $\pm$ sample standard deviation across the 15 test signals, with finite-SNR comparisons paired by using the same noise realisation for every method. Note that for display only, all T--F images are rendered after applying a $\sigma=1$-pixel isotropic Gaussian anti-aliasing filter to improve readability at publication scale; this display filter is not used in the quantitative metrics.

To quantify the influence of the physics-based reprojection objective on model performance, we conduct a post-convergence sensitivity analysis. Starting from the tonal checkpoint trained with the nominal physics weighting, three models are continued independently, with the physics-term weights in Eq.~\eqref{eq:loss} rescaled to $0\times$, $1\times$, and $10\times$ their nominal values. All runs are initialised using restarted Adam optimisers with increased learning rates and are continued to convergence under identical stopping conditions. Performance is subsequently evaluated on 50 held-out signals from the dataset generated according to Section~\ref{data}, using the native log-frequency grid. Predicted ITFR fidelity is assessed using the $BC$, $JS$, and $RER$ metrics, while physics reprojection fidelity is evaluated via $BC$, $JS$, and a channel-normalised relative Frobenius reprojection error defined as:
\begin{smallalign*}
&\mathcal{E}_{\mathrm{physics}, n} = \frac{1}{K}\sum_{k=1}^K\frac{ \|\mathbf{\tilde{\hat{y}}}_{n,k}-\mathbf{\tilde{y}}_{n,k}\|_F}{\|\mathbf{\tilde{y}}_{n,k}\|_F},\\
&\text{where:}\qquad\mathbf{\tilde{y}}_{n,k}=\frac{\mathbf{y}_{n,k}}{\sum_{t,f} y_{n,k, t, f}},
\qquad
\mathbf{\tilde{\hat{y}}}_{n,k}=\frac{\mathbf{\hat{y}}_{n,k}}{\sum_{t,f} \hat{y}_{n,k, t, f}}.
\end{smallalign*}
The physics-domain $\mathrm{BC}$ and $\mathrm{JS}$ values are likewise computed independently for each CLAWT channel and averaged equally over the $K$-channel constellation. The results are summarised in Table~\ref{tab:physics_ablation}, and Fig.~\ref{fig:example_physics} provides a qualitative illustration of the effect of varying the physics contribution.

Finally, we evaluate the tonal and harmonic PHAST-Net models and their spline counterparts on the same held-out set directly in the \textit{native log-frequency} domain to assess spline post-processing and fundamental isolation; the harmonic comparison uses the proposed Harmonic ITFR target, for which no directly comparable existing method was identified. The results are presented in Table~\ref{tab:native_harmonic}.

\vspace{-3mm}\section{Discussion}

\begin{figure*}
    \centering
    \vspace{-0mm}\includegraphics[
        width=1.0\textwidth,
        trim={0.3cm 0.25cm 0.21cm 0.7cm}, %LBRT
        clip
    ]{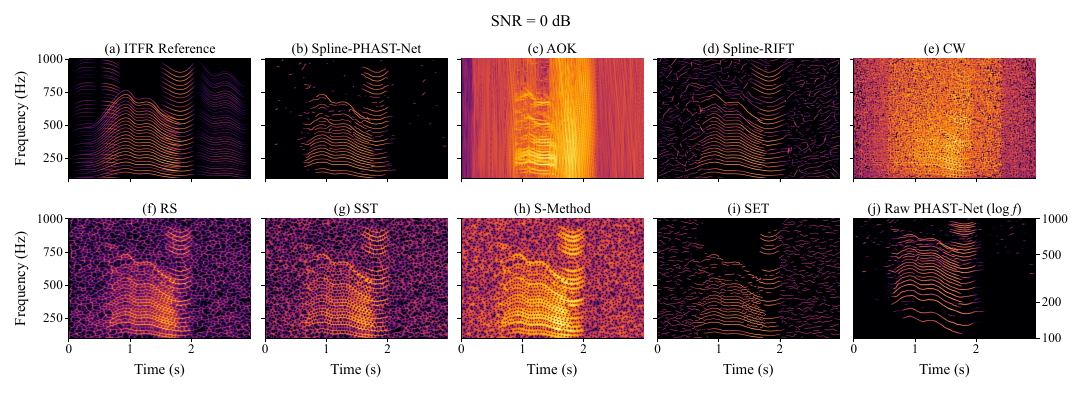}
    \vspace{-6mm}
    \caption{Example signal from the comparative evaluation
with additive white Gaussian noise (AWGN) at an SNR of \(0\,\mathrm{dB}\).}
    \label{fig:evaluation}\vspace{-2mm}
\end{figure*}

As observed in Table~\ref{tab:metrics_combined}, Spline-PHAST-Net achieves the best mean performance among the evaluated methods across the reported metrics and SNR conditions. In particular, Fig.~\ref{fig:evaluation} demonstrates that both the raw PHAST-Net and Spline-PHAST-Net representations substantially attenuate cross-term interference and noise relative to other approaches, yielding representations that more closely approach the reference ITFR. Moreover, while most methods exhibit difficulty in this setting, the proposed algorithm effectively accommodates the pronounced T--F-varying component curvature, overtone structure, close component spacing, and component intersections. 

\begin{figure*}
    \centering
    \vspace{-0mm}\includegraphics[
        width=1\textwidth,
        trim={0.3cm 0.45cm 0.7cm 0.7cm}, %LBRT
        clip
    ]{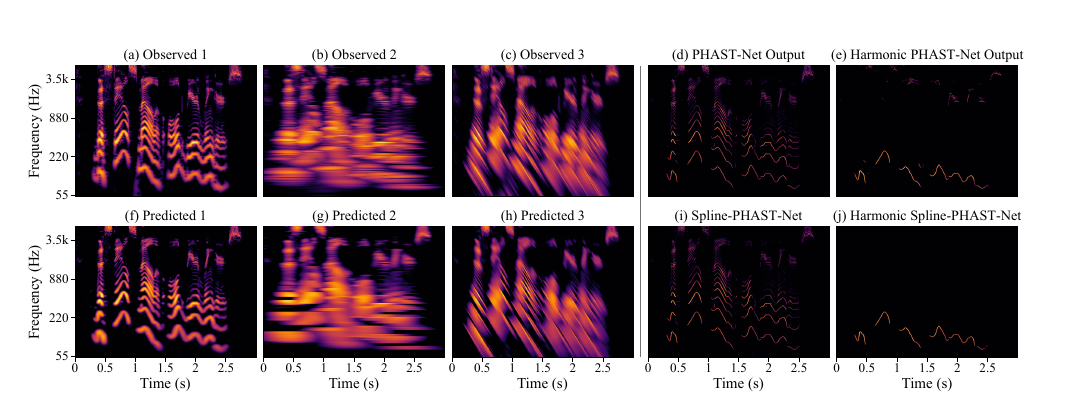}
    \vspace{-7mm}
    \caption{A visualisation of the PHAST-Net framework applied to a speech excerpt from the “Harvard sentences” database~\cite{speech}. On the left, the action of the physics operator is illustrated for several elements of the CLAWT constellation: panels (a)–(c) display the observed (input) CLAWTs, $\mathbf{\Phi}_k$, while panels (f)–(h) show the corresponding predicted, cross-term-free CLAWTs, $\mathbf{\hat{y}}_k=\left[\mathcal{P}(\mathbf{\hat{x}})\right]_k$, computed from the tonal output in panel (d). The specific constellation elements considered here are members $k=1$, 3, and 14. On the right, panels (d) and (e) present the PHAST-Net outputs, $\mathbf{\hat{x}}$, for the tonal and harmonic models, respectively, and panels (i) and (j) depict the associated spline-based representations. \vspace{-3mm}}
    \label{fig:evaluation_speech}
\end{figure*}

\begin{figure*}
    \centering
    \vspace{-0mm}\includegraphics[
        width=1\textwidth,
        trim={0.3cm 0.45cm 0.7cm 0.7cm}, %LBRT
        clip
    ]{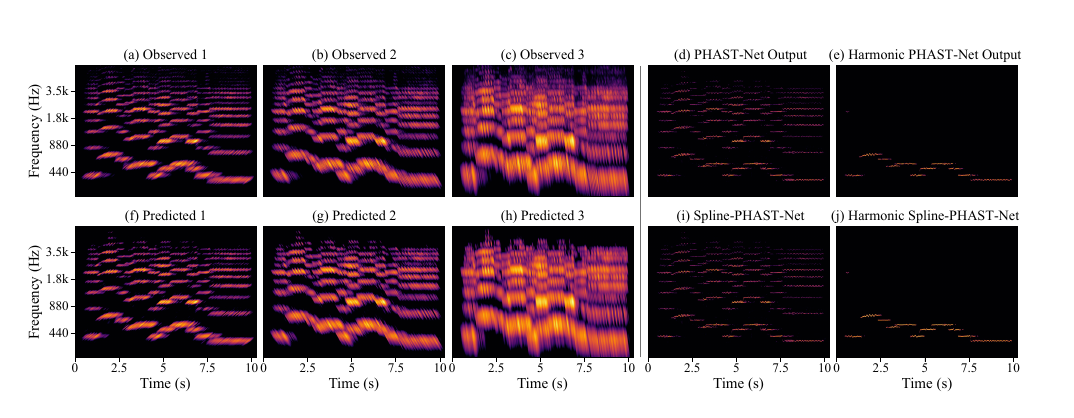}
    \vspace{-7mm}
    \caption{A visualisation of the PHAST-Net framework applied to a short violin extract from the piece Piano Concerto No. 4 by the author. As per the speech example in Fig.~\ref{fig:evaluation_speech}, several elements of the CLAWT constellation are shown, with panels (a)–(c) displaying the observed (input) CLAWTs, $\mathbf{\Phi}_k$, and panels (f)–(h) showing the corresponding predicted, cross-term-free CLAWTs, $\mathbf{\hat{y}}_k=\left[\mathcal{P}(\mathbf{\hat{x}})\right]_k$, computed from the tonal output in panel (d). The constellation elements considered here are members $k=9$, 11, and 14. On the right, panels (d) and (e) present the PHAST-Net outputs, $\mathbf{\hat{x}}$, for the tonal and harmonic models, respectively, and panels (i) and (j) depict the associated spline-based representations.\vspace{-3mm}}
    \label{fig:evaluation_music}
\end{figure*}

\begin{figure*}
    \centering
    \vspace{-0mm}\includegraphics[
        width=1\textwidth,
        trim={0.3cm 0.45cm 0.7cm 0.7cm}, %LBRT
        clip
    ]{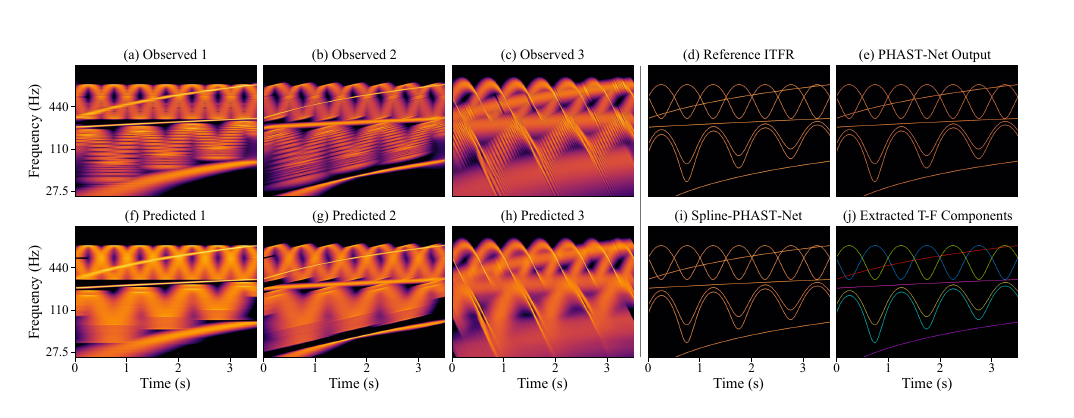}
    \vspace{-7mm}
    \caption{A visualisation of the framework applied to the evaluative signal $x_2(t)$ (Eq.~\eqref{eq:complex_signal_log} in the SM). As before, on the left, several elements of the CLAWT constellation are presented in panels (a)–(c) (the observed input CLAWTs, $\mathbf{\Phi}_k$), and panels (f)–(h) (the corresponding predicted, cross-term-free CLAWTs, $\mathbf{\hat{y}}_k=\left[\mathcal{P}(\mathbf{\hat{x}})\right]_k$). The specific constellation elements considered here are members $k=3$, 9, and 14. On the right, panels (d) and (e) present the Reference target ITFR, $\mathbf{x}$, and PHAST-Net output, $\mathbf{\hat{x}}$,  respectively, and panel (i) depicts the associated spline-based representation with the extracted T--F components (colour-coded) shown in (j).}
    \label{fig:evaluation_synthetic}
\end{figure*}

For the challenging high-complexity waveform defined in Eq.~\ref{eq:complex_signal_log} and evaluated in Fig.~\ref{fig:evaluation_synthetic}, the algorithm significantly reduces cross-term interference in the predicted CLAWT constellation, as evidenced by panels (a)–(c) relative to (f)–(h), while simultaneously preserving high instantaneous-frequency precision in the predicted ITFR. Similarly, the Spline-PHAST-Net algorithm reliably performs component extraction and data association, despite the intricate intersection behaviour present in the T--F plane. For the speech excerpt in Fig.~\ref{fig:evaluation_speech}, the algorithm again substantially suppresses cross-term interference, dynamically adapting to the complex T--F-varying curvature and mitigating interactions between harmonics. This yields high-precision IF ridge estimates and accurate fundamental-trajectory extraction, with panels (e) and (j) exhibiting strong harmonic suppression that exposes the underlying fundamental speech contour. The music excerpt in Fig.~\ref{fig:evaluation_music} likewise illustrates that the method accurately recovers the fundamental structure, at frequencies consistent with the musical score, while capturing locally prominent frequency-modulated vibrato. These results demonstrate the versatility of the approach and its robustness to variations in overtone structure and harmonic interaction. 

\begin{figure*}
    \centering
    \vspace{-0mm}\includegraphics[
        width=1.0\textwidth,
        trim={0.3cm 0.45cm 0.7cm 0.7cm}, %LBRT
        clip
    ]{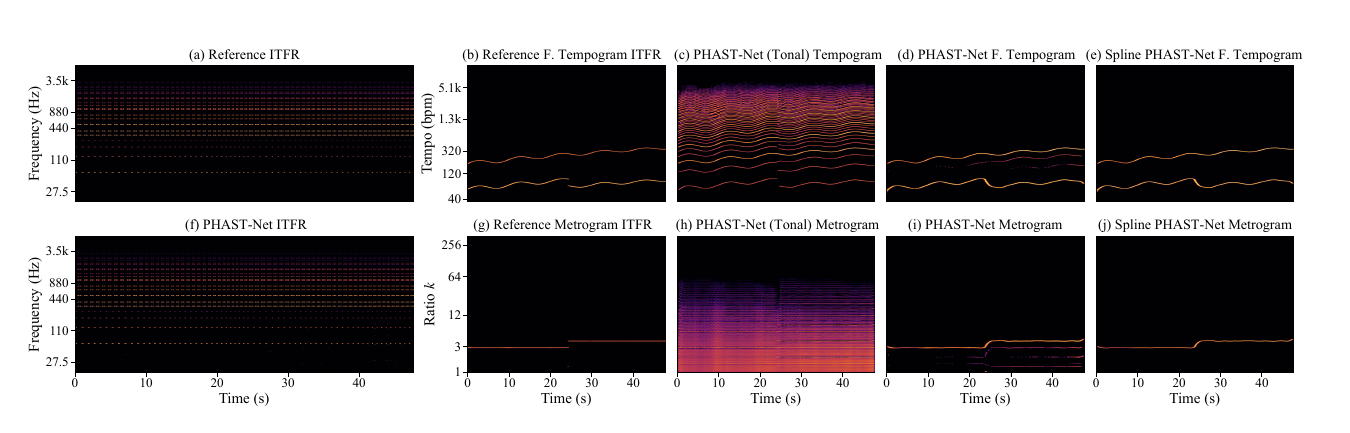}
    \vspace{-3mm}
    \caption{A visualisation of the Fundamental Tempogram and Metrogram inference process: (a) provides the reference target ITFR, $\mathbf{x}$, for the simulated musical excerpt, and (f) provides the raw predicted PHAST-Net output, $\mathbf{\hat{x}}$. Panels (b) and (g) present the reference Fundamental Tempogram and Metrogram ITFRs provided by Eq.~\eqref{ex:tempogram} and Eq.~\eqref{ex:metrogram}, respectively. Panels (c) and (d) present the Tempogram PHAST-Net outputs computed from the onset detection function generated from the PHAST-Net output, $\mathbf{\hat{x}}$ (panel (f)), employing the tonal and harmonic models, respectively, and panels (h) and (i) are the corresponding Metrograms, $\mathcal{M}(k,t)$. Finally, panels (e) and (j) are the Spline-PHAST-Net outputs computed from the representations in panels (d) and (i).}
    \label{fig:evaluation_tempo/metrogram}\vspace{-2.5mm}
\end{figure*}

Likewise, the PHAST-Net–inferred Fundamental Tempogram and Metrogram representations for the synthetic example in Fig.~\ref{fig:evaluation_tempo/metrogram} closely approximate the corresponding idealised representations, exhibiting negligible visible cross-term interference and high T--F resolution, while capturing the pronounced tempo fluctuations. In particular, as shown in panels (c) and (h) relative to (d) and (i), the employment of a harmonic model for Tempogram inference effectively suppresses overtone structure, thereby isolating the fundamental beat and bar-tempo trajectories (panel (d)) and producing a substantially improved Metrogram (panel (i)) with the expected two metric trajectories, corresponding to the 3/4-metre and 4/4-metre sections. 

\begin{figure*}
    \centering
    \vspace{-0mm}\includegraphics[
        width=1\textwidth,
        trim={0.3cm 0.45cm 0.7cm 0.5cm}, %LBRT
        clip
    ]{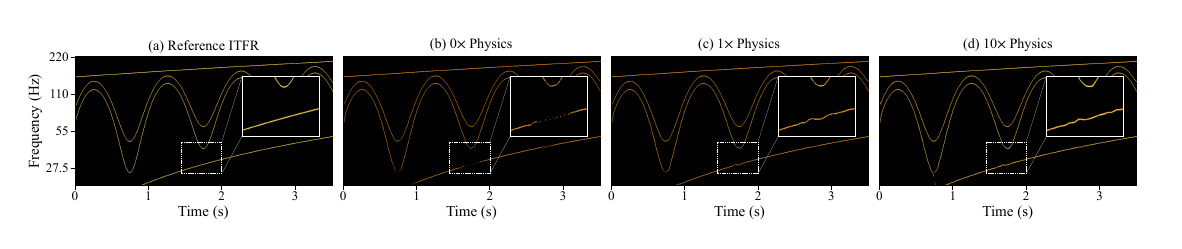}
    \vspace{-6mm}
    \caption{Post-convergence physics-loss sensitivity illustration for Eq.~\eqref{eq:physics}. (a) displays the reference target ITFR, $\mathbf{x}$, and (b)-(d) illustrate the PHAST-Net outputs, $\mathbf{\hat{x}}$, after continued optimisation with (b) $0\times$, (c) $1\times$, and (d) $10\times$ the nominal physics-loss weighting. The boxes indicate a challenging low-frequency cross-term-heavy region, with associated magnified insets highlighting the differences in local ridge continuity and distributional fidelity.\vspace{-3mm}}
    \label{fig:example_physics}
\end{figure*}

In the physics sensitivity study, Table~\ref{tab:physics_ablation} indicates that incorporating the physics-based reprojection term yields improvements in the $\mathrm{BC}$ and $\mathrm{JS}$ metrics in both the native ITFR and the physics-reprojected domains, alongside reductions in reprojection error, consistent with improved agreement between normalised energy distributions under the physics operator. For the nominal $\times 1$ model, this improvement in distributional fidelity is observed for BC and JS at all tested tolerances $\sigma_{\mathrm{eval}}$ except $\sigma_{\mathrm{eval}}=1$, with RER exhibiting a small reduction (Table~\ref{tab:physics_sigma_sensitivity}). Notably, the $10\times$ configuration attains the strongest distributional performance among the evaluated models. Consequently, these results indicate a trade-off between distributional fidelity and ridge concentration, consistent with the empirical choice of the intermediate nominal weighting used in PHAST-Net. In particular, the $1\times$ setting corresponds to a relative scaling of $\lambda_{\mathrm{physics}}=0.85\,\lambda_{\mathrm{image}}$ in Eq.~\eqref{eq:loss}, where these denote the common scalings of the physics- and image-domain loss terms, respectively. This behaviour is illustrated in Fig.~\ref{fig:example_physics}: the highlighted regions show missing T--F ridge structure for the $0\times$ model where the input exhibits strong cross-term interference, whereas the corresponding T--F content is retained in both physics-weighted models.

Finally, in the native tonal and harmonic PHAST-Net evaluation, the tonal PHAST-Net model and its associated spline variant considerably improve distributional T--F fidelity and ridge concentration relative to the input isotropic CLAWT. Likewise, the harmonic models substantially outperform the tonal model in harmonic suppression, indicating successful generalisation on held-out examples of the complex inhomogeneous Harmonic ITFR target.

\vspace{-0mm}

\vspace{-0mm}

\vspace{-0mm}

\vspace{-0mm}

\vspace{-0mm}

\vspace{-0mm}

\begin{table*}[t]
\caption{Evaluation metrics for the comparative benchmark. Cells report the mean $\pm$ sample standard deviation across signal instances. Bold indicates the best mean value per row.}
\label{tab:metrics_combined}
\centering
\footnotesize
\setlength{\tabcolsep}{4pt}
\vspace{-2mm}\begin{tabular}{l|cccccccc}
\hline
\multicolumn{9}{l}{\textbf{Bhattacharyya overlap ($\uparrow$)}}\\
\cline{1-9}
SNR (dB) & AOK & Choi-Williams & Reassignment & S-Method & SET & SST & Spline-PHAST-Net & Spline-RIFT \\ \hline
-5 dB & 0.511 $\pm$ 0.068 & 0.530 $\pm$ 0.044 & 0.716 $\pm$ 0.036 & 0.693 $\pm$ 0.017 & 0.693 $\pm$ 0.067 & 0.739 $\pm$ 0.026 & \textbf{0.787 $\pm$ 0.059} & 0.761 $\pm$ 0.070 \\
0 dB & 0.530 $\pm$ 0.060 & 0.587 $\pm$ 0.041 & 0.804 $\pm$ 0.051 & 0.759 $\pm$ 0.012 & 0.770 $\pm$ 0.074 & 0.823 $\pm$ 0.037 & \textbf{0.879 $\pm$ 0.045} & 0.831 $\pm$ 0.059 \\
5 dB & 0.528 $\pm$ 0.060 & 0.625 $\pm$ 0.039 & 0.842 $\pm$ 0.058 & 0.787 $\pm$ 0.011 & 0.799 $\pm$ 0.076 & 0.860 $\pm$ 0.043 & \textbf{0.913 $\pm$ 0.038} & 0.854 $\pm$ 0.054 \\
$\infty$ & 0.521 $\pm$ 0.060 & 0.669 $\pm$ 0.039 & 0.862 $\pm$ 0.062 & 0.801 $\pm$ 0.010 & 0.812 $\pm$ 0.076 & 0.879 $\pm$ 0.046 & \textbf{0.929 $\pm$ 0.035} & 0.861 $\pm$ 0.051 \\
\hline\hline
\multicolumn{9}{l}{\textbf{Jensen--Shannon divergence ($\downarrow$)}}\\
\cline{1-9}
SNR (dB) & AOK & Choi-Williams & Reassignment & S-Method & SET & SST & Spline-PHAST-Net & Spline-RIFT \\ \hline
-5 dB & 0.394 $\pm$ 0.047 & 0.377 $\pm$ 0.031 & 0.234 $\pm$ 0.032 & 0.253 $\pm$ 0.011 & 0.240 $\pm$ 0.052 & 0.215 $\pm$ 0.023 & \textbf{0.168 $\pm$ 0.046} & 0.189 $\pm$ 0.054 \\
0 dB & 0.381 $\pm$ 0.042 & 0.334 $\pm$ 0.028 & 0.167 $\pm$ 0.045 & 0.203 $\pm$ 0.007 & 0.183 $\pm$ 0.059 & 0.150 $\pm$ 0.033 & \textbf{0.098 $\pm$ 0.035} & 0.139 $\pm$ 0.047 \\
5 dB & 0.382 $\pm$ 0.041 & 0.305 $\pm$ 0.027 & 0.138 $\pm$ 0.051 & 0.182 $\pm$ 0.006 & 0.161 $\pm$ 0.061 & 0.122 $\pm$ 0.038 & \textbf{0.071 $\pm$ 0.031} & 0.121 $\pm$ 0.043 \\
$\infty$ & 0.387 $\pm$ 0.042 & 0.272 $\pm$ 0.027 & 0.122 $\pm$ 0.054 & 0.171 $\pm$ 0.005 & 0.151 $\pm$ 0.062 & 0.107 $\pm$ 0.041 & \textbf{0.058 $\pm$ 0.028} & 0.116 $\pm$ 0.041 \\
\hline\hline
\multicolumn{9}{l}{\textbf{RER ($\uparrow$)}}\\
\cline{1-9}
SNR (dB) & AOK & Choi-Williams & Reassignment & S-Method & SET & SST & Spline-PHAST-Net & Spline-RIFT \\ \hline
-5 dB & 0.338 $\pm$ 0.075 & 0.337 $\pm$ 0.070 & 0.556 $\pm$ 0.064 & 0.417 $\pm$ 0.049 & 0.679 $\pm$ 0.065 & 0.545 $\pm$ 0.056 & \textbf{0.823 $\pm$ 0.049} & 0.778 $\pm$ 0.050 \\
0 dB & 0.356 $\pm$ 0.079 & 0.374 $\pm$ 0.069 & 0.650 $\pm$ 0.094 & 0.457 $\pm$ 0.049 & 0.773 $\pm$ 0.083 & 0.636 $\pm$ 0.082 & \textbf{0.903 $\pm$ 0.038} & 0.863 $\pm$ 0.040 \\
5 dB & 0.360 $\pm$ 0.082 & 0.404 $\pm$ 0.069 & 0.695 $\pm$ 0.111 & 0.475 $\pm$ 0.050 & 0.810 $\pm$ 0.089 & 0.680 $\pm$ 0.098 & \textbf{0.930 $\pm$ 0.033} & 0.891 $\pm$ 0.034 \\
$\infty$ & 0.358 $\pm$ 0.082 & 0.448 $\pm$ 0.072 & 0.722 $\pm$ 0.123 & 0.485 $\pm$ 0.051 & 0.828 $\pm$ 0.093 & 0.705 $\pm$ 0.108 & \textbf{0.945 $\pm$ 0.031} & 0.902 $\pm$ 0.031 \\
\hline
\end{tabular}\vspace{-3mm}
\end{table*}

\begin{table*}[t]
\centering
\caption{Post-convergence physics-loss sensitivity study. Cells report mean $\pm$ sample standard deviation over 50 signals. 0$\times$, 1$\times$, and 10$\times$ correspond to the relative physics-loss weighting. Physics reprojection metrics are averaged equally over the 15 constellation channels. Bold indicates the best mean value within each domain and metric.}
\label{tab:physics_ablation}
\footnotesize
\setlength{\tabcolsep}{4pt}
\vspace{-2mm}\begin{tabular}{l|ccc|ccc}
\hline
& \multicolumn{3}{c|}{\textbf{ITFR-domain fidelity} ($\mathbf{\hat{x}}$)} & \multicolumn{3}{c}{\textbf{Physics reprojection fidelity} ($\mathbf{\hat{y}}=\mathcal{P}(\mathbf{\hat{x}})$)} \\
\cline{2-7}
Metric & 0$\times$ & 1$\times$ & 10$\times$ & 0$\times$ & 1$\times$ & 10$\times$ \\ \hline
Bhattacharyya overlap ($\uparrow$) & 0.9625 $\pm$ 0.0459 & 0.9636 $\pm$ 0.0456 & \textbf{0.9651 $\pm$ 0.0437} & 0.9857 $\pm$ 0.0152 & 0.9902 $\pm$ 0.0087 & \textbf{0.9908 $\pm$ 0.0082} \\
Jensen--Shannon divergence ($\downarrow$) & 0.0319 $\pm$ 0.0375 & 0.0311 $\pm$ 0.0375 & \textbf{0.0299 $\pm$ 0.0359} & 0.0132 $\pm$ 0.0136 & 0.0091 $\pm$ 0.0079 & \textbf{0.0087 $\pm$ 0.0075} \\
RER ($\uparrow$) / relative Frobenius error ($\downarrow$) & \textbf{0.9622 $\pm$ 0.0479} & 0.9492 $\pm$ 0.0656 & 0.9469 $\pm$ 0.0654 & 0.2218 $\pm$ 0.1212 & 0.1722 $\pm$ 0.0778 & \textbf{0.1655 $\pm$ 0.0812} \\
\hline\hline
\end{tabular}\vspace{-3mm}
\end{table*}

\begin{table*}[t]
\caption{Native tonal and harmonic PHAST-Net evaluation. Cells report mean $\pm$ sample standard deviation over the 50 held-out signals. Bold indicates the best mean value within each target and metric.}
\label{tab:native_harmonic}
\centering\footnotesize
\setlength{\tabcolsep}{4pt}
\vspace{-2mm}\begin{tabular}{l|ccc|ccc}
\hline
& \multicolumn{3}{c|}{\textbf{ITFR target}} & \multicolumn{3}{c}{\textbf{Harmonic ITFR target}} \\
\cline{2-7}
Method & BC $\uparrow$ & JS $\downarrow$ & RER $\uparrow$ & BC $\uparrow$ & JS $\downarrow$ & RER $\uparrow$ \\ \hline
Isotropic CLAWT & 0.8394 $\pm$ 0.0748 & 0.1327 $\pm$ 0.0598 & 0.6167 $\pm$ 0.1329 & 0.5136 $\pm$ 0.1455 & 0.3754 $\pm$ 0.1090 & 0.2867 $\pm$ 0.1685 \\
PHAST-Net & \textbf{0.9639 $\pm$ 0.0457} & \textbf{0.0307 $\pm$ 0.0373} & \textbf{0.9506 $\pm$ 0.0634} & 0.6095 $\pm$ 0.1430 & 0.2977 $\pm$ 0.1075 & 0.4690 $\pm$ 0.1890 \\
Spline-PHAST-Net & 0.9622 $\pm$ 0.0459 & 0.0322 $\pm$ 0.0374 & 0.9499 $\pm$ 0.0634 & 0.6081 $\pm$ 0.1442 & 0.2988 $\pm$ 0.1083 & 0.4691 $\pm$ 0.1905 \\
Harmonic PHAST-Net & -- & -- & -- & \textbf{0.8465 $\pm$ 0.0958} & \textbf{0.1281 $\pm$ 0.0758} & 0.9081 $\pm$ 0.0829 \\
Harmonic Spline-PHAST-Net & -- & -- & -- & 0.8440 $\pm$ 0.0958 & 0.1300 $\pm$ 0.0755 & \textbf{0.9090 $\pm$ 0.0818} \\
\hline\hline
\end{tabular}\vspace{-4mm}
\end{table*}

\vspace{-4mm} \section{Conclusion}
In conclusion, PHAST-Net establishes a unified, physics-informed framework for estimating high-resolution Ideal Time--Frequency Representations across spectral, tempo-based, metrical, and harmonic domains. By combining a strategically selected CLAWT constellation derived from WVD-kernel analysis, attention-based cross-term suppression, and an auxiliary physics-informed reconstruction loss that promotes transform-domain consistency, the proposed approach improves representation accuracy, robustness to noise, and interpretability for nonstationary signal analysis. Harmonic PHAST-Net and Spline-PHAST-Net extend the framework by enabling fundamental-structure isolation, component-level extraction, and arbitrary-grid re-rendering. These results demonstrate that PHAST-Net provides a principled and generalisable foundation for advanced T--F analysis across speech, music, and related nonstationary domains.

\ifSMonly
    \setcounter{page}{1}
\fi

\clearpage

\section*{Supplementary Material}
\renewcommand{\thesubsection}{\Alph{subsection}}

\vspace{-0mm} \subsection{Harmonic ITFR properties}
\label{Harmonic_ITFR_properties}

Note that the following relationship holds between Eq.~\eqref{ITFR__h_Eq2} and Eq.~\eqref{Harmonic_ITFR_log}:
\begin{smallalign}
    &\mathrm{ITFR}_{s}\!\{z_{H}\}\!\left(s, t\right) \notag\\&\quad= \!\sum_{p=1}^P \!\sum_{n_h=1}^{N_{H,p}} \!\!\mathds{1}_{{\cal \tau}_p}\!\left(t\right) \!A_h(n_h, p, t) A_p(t) \delta\left(s \!- \!s_p(t) \!- \!s_h(n_h)\right) \label{harmonic invariance}\\
    &\quad=  \sum_{p=1}^P\!\int_{\mathbb{R}}\mathrm{ITFR}_{H,s}\{z_{H,\,p}(t)\}\!\left(s \!- \!s', t\right) \,H_p\!\left(s', t\right)\, d s'\\
    &\quad=  \!\sum_{p=1}^P\left[\mathrm{ITFR}_{H,s}\{z_{H, \, p}(t)\} *_s H_p\right]\left(s, \, t\right),
\end{smallalign}
\noindent where $*_s$ denotes convolution with respect to $s$, $s_h(n_h)=12 \log_2\!{n_h}$, and:
\begin{smallalign}
\label{Harmonic_itfr_log}
    H_p\left(s, \, t\right) &= \sum_{n_h=1}^{N_{H,p}} A_h(n_h, p, t) \delta\left(s - s_h(n_h)\right),
\end{smallalign}
\noindent where Eq.~\eqref{Harmonic_itfr_log} is the inhomogeneous (T--F varying) Harmonic distribution. Thus, crucially, by reformulating the Harmonic ITFR and the ITFR in the logarithmic domain, the relationship becomes convolutional, given that the relative positions of the harmonics become T--F invariant as observed in Eq.~\eqref{harmonic invariance}. If we now let $z_{\bar{H}}\left(t\right)$ be a multi-component harmonic complex signal with a T--F \textit{invariant} (homogeneous) linearly distributed harmonic distribution:
\begin{smallalign}
\label{signal_bar_H}
    z_{\bar{H}}\!\left(t\right) \!&=
    \!\sum_{p=1}^P \!\sum_{n_h=1}^{N_h} \mathds{1}_{{\cal \tau}_p}\!\left(t\right)A_h(n_h) A_p(t)\exp\!{\left(j \!\left[\phi_{p,n_h, 1} \!+ \!n_h \!\cdot \!\phi_{p}\left(t\right)\right]\right)}.
\end{smallalign}
\noindent where $A_h(n_h)$ is the relative amplitude of the $n_h$th ($1 \le n_h \le N_h$) harmonic ($\sum_{n_h}A_h(n_h)=1$). As before, a suitable definition for this \textit{linear} $\omega$ domain Homogeneous Harmonic ITFR, $\mathrm{ITFR}_{\bar{H}, \omega}$, for signal $z_{\bar{H}}\left(t\right)$ is:
\begin{smallalign}
\label{ITFR_barh_full_Eq}
\mathrm{ITFR}_{\bar{H}, \omega}\{z_{\bar{H}}\}\left(\omega, \, t\right) &= \sum_{p=1}^P \mathds{1}_{{\cal \tau}_p}\!\left(t\right)A_p(t)\delta\left(\omega - \omega_p(t)\right).
\end{smallalign}
Note that the linear $\omega$ domain ITFR, $\mathrm{ITFR}_{\omega}$ (Eq.~\eqref{ITFR_linear}), for signal $z_{\bar{H}}\left(t\right)$ would likewise be defined as:
\begin{smallalign}
\label{ITFR__h_Eq}
    \mathrm{ITFR}_{\omega}\{z_{\bar{H}}\}\!\left(\omega, t\right) = \!\sum_{p=1}^P \!\sum_{n_h=1}^{N_h} \!\!\mathds{1}_{{\cal \tau}_p}\!\!\left(t\right)\! A_h(n_h) A_p(t) \delta\!\left(\omega \!- n_h\omega_p(t)\right).
\end{smallalign}
Expressing the ITFR for this homogeneous harmonic model in the semitonal domain, $\mathrm{ITFR}_{s}$, using the same change of coordinate system $\omega \mapsto s$ derived in Eq.~\eqref{ITFR_Eq_linear} provides:
\begin{smallalign}
    &\mathrm{ITFR}_{s}\!\{z_{\bar{H}}\}\!\left(s, t\right) \notag \\&= \!\sum_{p=1}^P \!\sum_{n_h=1}^{N_h} \!\!\mathds{1}_{{\cal \tau}_p}\!\left(t\right) \!A_h(n_h) A_p(t) \delta\left(s \!- \!s_p(t) \!- \!s_h(n_h)\right) \\
    &=  \!\int_{\mathbb{R}}\mathrm{ITFR}_{\bar{H},s}\{z_{\bar{H}}(t)\}\!\left(s \!- \!s', t\right) \,\bar{H}\!\left(s'\right)\, d s'\\
    &=  \left[\mathrm{ITFR}_{\bar{H},s}\{z_{\bar{H}}(t)\} *_s \bar{H}\right]\left(s, \, t\right),
\end{smallalign}
\noindent with:
\begin{smallalign}
    \mathrm{ITFR}_{\bar{H},s}\{z_{\bar{H}}\} &= \sum_{p=1}^P \mathds{1}_{{\cal \tau}_p}\!\left(t\right)A_p(t)\delta\left(s-s_p(t)\right) \label{Harmonic_Hom_ITFR_log} \\
    \bar{H}\left(s\right) &= \sum_{n_h=1}^{N_h} A_h(n_h) \delta\left(s - s_h(n_h)\right) \label{barH},
\end{smallalign}
\noindent where Eq.~\eqref{Harmonic_Hom_ITFR_log} is the homogeneous Harmonic ITFR, and Eq.~\eqref{barH} is the Homogeneous Harmonic distribution. Thus, in the homogeneous case, the ITFR can be expressed as a single convolution between the Harmonic ITFR and a Harmonic distribution (Eq.~\eqref{barH}).

\vspace{-2mm} \subsection{Derivation of the Metrogram ITFR for an idealised Metric extract}

\label{metrogram_derivation}

To simulate a simple music extract transitioning from 3/4 to 4/4, let the idealised input signal, $z_{M}(t)$, be a chain of regularly repeating block chords (time-invariant amplitudes) corresponding to both upbeats and downbeats such that:
\begin{smallalign}
z_M(t) \!&= \!\sum_{n=1}^\infty z_{beat}\left(t-\phi_{beat}^{-1}(2 \pi n)\right)\! +\! \sum_{m=1}^\infty z_{bar}\left(t-\phi_{bar}^{-1}(2 \pi m)\right),
\end{smallalign}
\noindent where $z_{beat}(t)$ and $z_{bar}(t)$ are derived from $z_H(t)$ in Eq.~\eqref{signal_harmonic_ITFR} with $\mathds{1}_{{\cal \tau}_p} = \mathds{1}_{beat}$ and $\mathds{1}_{{\cal \tau}_p} = \mathds{1}_{bar}$, $A_h(n_h, p, t)=A_{h, beat}(n_h, p)$ and $A_h(n_h, p, t)=A_{h, bar}(n_h, p)$, $P=P_{beat}$ and $P=P_{bar}$, and $A_p(t) = A_{p, beat}$ and $A_p(t) = A_{p, bar}$ respectively. Note that $\mathcal{\tau}_{beat}=[0,T_{beat}]$ and $\mathcal{\tau}_{bar}=[0,T_{bar}]$ for the indicator function. Likewise, $\phi_{beat}^{-1}(x)$ and $\phi_{bar}^{-1}(x)$ are the inverse functions for the beat and downbeat phase, such that:
\begin{smallalign}
\label{phase_tempo}
\phi_{beat}(t) &= \phi_{beat, 0}+\int_0^t \omega_{beat}(\tau) \, d \tau, \\
\phi_{bar}(t) &= \frac{\phi_{beat, 0}}{k(0)} +\int_0^t \frac{\omega_{beat}(\tau)}{k(\tau)} \, d \tau,
\end{smallalign}
\noindent where $\phi_{beat, 0}$ is the initial phase at $t=0$, $\omega_{beat}(t)$ is the time-varying angular beat frequency (with $\omega_{beat}(t)=\omega_{bar}(t) \cdot k(t)$), and $k(t)$ is the time-dependent time signature (metric) numerator. For the simple case of a transition from $N_{3/4}$ bars of 3/4 (so $3 \times N_{3/4}$ beats) to 4/4, $k(t)$ can be defined as such:
\begin{smallalign}
k(t) &= 
\begin{cases} 
3 & \quad 0 \le t \le \phi_{beat}^{-1}(2 \pi \cdot 3 \cdot N_{3/4}) \\
4 & \quad t > \phi_{beat}^{-1}(2 \pi \cdot 3 \cdot N_{3/4})
\end{cases}.
\end{smallalign}
Thus:
\begin{smallalign}
&f_{\text{onset}}\left\{z_M\right\}(t) \\
&= \left(\frac{\partial}{\partial t}\int_{\mathbb{R}}\mathrm{ITFR}_{H, s}\{z_M\}(s,t)\,ds\right)_+ \\
&= \left(\frac{\partial}{\partial t}\int_{0}^{\infty}\mathrm{ITFR}_{H, \omega}\{z_M\}(\omega,t)\,d\omega\right)_+ \\
&= \left(\frac{\partial}{\partial t}\sum_{n=1}^\infty \mathds{1}_{beat}\!\left(t-\phi_{beat}^{-1}(2 \pi n)\right)\left(\sum_{p=1}^{P_{beat}} A_{p, beat}\right) \right.\\
&\hspace{15mm} \left.+ \sum_{m=1}^\infty \mathds{1}_{bar}\!\left(t-\phi_{bar}^{-1}(2 \pi m)\right)\left(\sum_{p=1}^{P_{bar}} A_{p, bar}\right)\right)_+.
\end{smallalign}
Let $A_{beat} = \sum_{p=1}^{P_{beat}} A_{p, beat}$ and $A_{bar} = \sum_{p=1}^{P_{bar}} A_{p, bar}$. Thus:
\begin{smallalign}
&f_{\text{onset}}\left\{z_M\right\}(t) \notag\\
&= \sum_{n=1}^\infty A_{beat} \, \delta\!\left(t-\phi_{beat}^{-1}(2 \pi n)\right) + \sum_{m=1}^\infty A_{bar} \, \delta\!\left(t-\phi_{bar}^{-1}(2 \pi m)\right).
\end{smallalign}
Therefore, the Fundamental Tempogram ITFR (Eq.~\eqref{fundamental_tempogram}) can be derived:
\begin{smallalign}
\text{T}_{F, \lambda}\left\{z_M\right\}\!(\lambda, \, t)&=\!\mathrm{ITFR}_{H,s}\left\{f_{\text{onset}}\left\{z_M\right\}(t)\right\}(\lambda, t) \notag\\
&\propto  
\!A_{beat} \,\delta\!\left(\lambda-\lambda_{beat}(t)\right)+A_{bar} \,\delta\!\left(\lambda-\lambda_{bar}(t)\right),
\end{smallalign}
\noindent where $\lambda_{beat}(t) = \omega_{s}^{-1}(\omega_{beat}(t))$ and $\lambda_{bar}(t) = \omega_{s}^{-1}(\omega_{bar}(t))$, and $\omega_{s}^{-1}(\omega)$ is the inverse function of Eq.~\eqref{semi}. Note that $\omega_{beat}(t)=\omega_{bar}(t) \cdot k(t)$, thus:
\begin{smallalign}
\lambda_{beat}(t) &= 12\log_2 \left(\frac{\omega_{bar}(t) \cdot k(t)}{2\pi f_0}\right) \\&=\lambda_{bar}(t) + 12\log_2 k(t),
\end{smallalign}
\noindent and:
\begin{smallalign}
\text{T}_{F, \lambda}\left\{z_M\right\}\!(\lambda, \, t) &\propto  \!A_{beat} \,\delta\!\left(\lambda-\lambda_{bar}(t) - 12\log_2 k(t)\right)\notag\\
&\hspace{15mm}+A_{bar} \,\delta\!\left(\lambda-\lambda_{bar}(t)\right).
\end{smallalign}
Therefore, the Metrogram ITFR can be derived
(Eq.~\eqref{metrogram_definition}):
\begin{smallalign}
\mathcal{M}(k, t)
&= \int_{\mathbb{R}}\text{T}_{F,\lambda}(\lambda, t)\,
\text{T}_{F,\lambda}(\lambda + 12\log_2 k, t)\, d\lambda \\
&\propto A_{beat} A_{bar} \tfrac{k(t) \ln{2}}{12}\, \delta\left(k-k(t)\right)
\\&\propto \begin{cases} 
\delta\left(k-3\right) & \quad 0 \le t \le \phi_{beat}^{-1}(2 \pi \cdot 3 \cdot N_{3/4}) \\
\delta\left(k-4\right) & \quad t > \phi_{beat}^{-1}(2 \pi \cdot 3 \cdot N_{3/4})
\end{cases},
\end{smallalign}
\noindent as desired. Note that $\mathcal{M}(k, t)$ here is restricted to the domain $k>1$.

\vspace{-0mm}

\begin{table*}[t]
\caption{Ranking stability across $\sigma_{\mathrm{eval}}$. For each $\sigma_{\mathrm{eval}}$, metrics are averaged across signals and SNRs. Each metric is min--max normalised across methods, with Jensen--Shannon divergence inverted since lower is better; the three normalised metrics are then averaged. Bold indicates the best method per $\sigma_{\mathrm{eval}}$.}
\label{tab:sigma_ablation}
\centering
\footnotesize
\setlength{\tabcolsep}{4pt}
\vspace{-2mm}\begin{tabular}{c|cccccccccc}
\hline
$\sigma$ & AOK & Choi-Williams & RIFT & Reassignment & S-Method & SET & SST & Spline-PHAST-Net & Spline-RIFT & WVD \\
\hline
1 & 0.005 & 0.151 & 0.740 & 0.613 & 0.350 & 0.638 & 0.623 & \textbf{1.000} & 0.829 & 0.010 \\
2 & 0.020 & 0.190 & 0.836 & 0.713 & 0.512 & 0.727 & 0.742 & \textbf{1.000} & 0.880 & 0.000 \\
3 & 0.048 & 0.215 & 0.870 & 0.787 & 0.641 & 0.781 & 0.818 & \textbf{1.000} & 0.895 & 0.000 \\
4 & 0.073 & 0.229 & 0.886 & 0.832 & 0.723 & 0.818 & 0.857 & \textbf{1.000} & 0.903 & 0.000 \\
5 & 0.095 & 0.236 & 0.894 & 0.858 & 0.770 & 0.843 & 0.874 & \textbf{1.000} & 0.909 & 0.000 \\
\hline
\end{tabular}\end{table*}

\begin{table}[t]
\caption{Sensitivity of the reported nominal ($1\times$) physics model performance to the ridge tolerance parameter $\sigma_{\mathrm{eval}}$. Positive values indicate improvement of the 1$\times$ physics model relative to the 0$\times$ model; the JS difference is therefore sign-reversed.}
\label{tab:physics_sigma_sensitivity}
\centering\footnotesize
\setlength{\tabcolsep}{5pt}
\vspace{-2mm}\begin{tabular}{c|ccc}
\hline
$\sigma_{\mathrm{eval}}$ & $\Delta$BC $\uparrow$ & $\Delta$JS $\uparrow$ & $\Delta$RER $\uparrow$ \\ \hline
1 & -0.0012 $\pm$ 0.0087 & -0.0009 $\pm$ 0.0066 & -0.0201 $\pm$ 0.0301 \\
2 & +0.0010 $\pm$ 0.0048 & +0.0008 $\pm$ 0.0039 & -0.0130 $\pm$ 0.0223 \\
3 & +0.0028 $\pm$ 0.0059 & +0.0022 $\pm$ 0.0048 & -0.0090 $\pm$ 0.0162 \\
4 & +0.0038 $\pm$ 0.0073 & +0.0032 $\pm$ 0.0060 & -0.0065 $\pm$ 0.0119 \\
5 & +0.0043 $\pm$ 0.0081 & +0.0037 $\pm$ 0.0069 & -0.0048 $\pm$ 0.0089 \\
\hline
\end{tabular}
\end{table}

\vspace{-2mm} \subsection{Derivation of the CLAWT kernel}
\label{derivation_kernel}

For an analysis function $W_{\omega}\!\left(t\right)$, the Wigner convolution identity provides \cite{cozens2025rift}:
\begin{smallalign}
\label{wvd_relation_harmonic_fractional}
    \Phi\!\left(\omega, t\right)
        &=2\pi\!\int_{\mathbb{R}^2}
            WVD_{z}\!\left(\omega_{1},T\right)\,
            WVD_{W_{\omega}}\!\left(\omega_{1},t-T\right)\,
            dT\,d\omega_{1}
\end{smallalign}

Now let the proposed wavelet function, $W_{\omega}\!\left(t\right)$, be newly defined as $W_{\sigma, \theta}^{(s)}\!\left(t\right) \triangleq A(t)e^{j\phi(t)}$, parameterised  by $(\sigma, \theta)$, such that: 
\begin{smallalign}
  A(t)\!=\!\dfrac{e^{-t^{2}/(2\sigma_s^{2})}}{\sqrt[4]{\pi\sigma_s^{2}}},
\;
  \phi(t)\!=\!\dfrac{2\pi f_{0}\,2^{s/12}}{\gamma}
          \left(e^{\gamma t}\!-\!1\right),\, s\!=\!\omega_s^{-1}(\omega),
\end{smallalign}
\noindent where $\gamma$ and $\sigma_s$ are defined in Eq.~\eqref{component_wavelet}. Thus, the WVD is:
\begin{smallalign}
  WVD_{W_{\sigma, \theta}^{(s)}}(\omega,t)&= \frac{1}{2 \pi}
  \int_{\mathbb{R}}
      A\!\left(t+\frac{\tau}{2}\right)A\!\left(t-\frac{\tau}{2}\right)
      \notag \\& \hspace{5mm} \cdot \exp{\left[j\,(\phi(t+\tau/2)-\phi(t-\tau/2))\right]}
      e^{-j\omega\tau}\,d\tau.    
\end{smallalign}
\noindent where:
\begin{smallalign*}
 A\!\left(t+\frac{\tau}{2}\right)A\!\left(t-\frac{\tau}{2}\right)
 =\frac{1}{\sqrt{\pi}\sigma_s}\,
   \exp\!\left[-\frac{t^{2}}{\sigma_s^{2}}
               -\frac{\tau^{2}}{4\sigma_s^{2}}\right],
\end{smallalign*}
\noindent and:
\begin{smallalign*}
   \phi\!\left(t+\frac{\tau}{2}\right)-\phi\!\left(t-\frac{\tau}{2}\right)
      &=\dfrac{\omega_0}{\gamma}
          \left(e^{\gamma \left(t + \frac{\tau}{2}\right)}-1\right) - \dfrac{\omega_0}{\gamma}
          \left(e^{\gamma \left(t - \frac{\tau}{2}\right)}-1\right) \\
      &=2\dfrac{\omega_0}{\gamma}e^{\gamma t}\left[
          \frac{e^{ \frac{\gamma \tau}{2}} - 
          e^{- \frac{\gamma \tau}{2}}}{2}\right] \\
      &=\frac{\omega_{0}}{\gamma}e^{\gamma t}\,
        2\sinh\!\left(\tfrac{\gamma\tau}{2}\right)
\end{smallalign*}
\noindent where $\omega_0 = 2\pi f_{0}\,2^{s/12}$. Thus:
\begin{smallalign}
\label{exact_WVD}
   &WVD_{W_{\sigma, \theta}^{(s)}}(\omega,t) \notag\\
   &\hspace{2mm}=\frac{1}{2 \pi}\frac{e^{-t^{2}/\sigma_s^{2}}}{\sqrt{\pi}\sigma_s}
    \int_{\mathbb{R}}
      \exp\!\left[-\frac{\tau^{2}}{4\sigma_s^{2}}
          -j\omega \tau + j\frac{\omega_{0}}{\gamma}e^{\gamma t}\,
        2\sinh\!\left(\tfrac{\gamma\tau}{2}\right)\right]\,d\tau.
  \end{smallalign}
Substituting Eq.~\eqref{exact_WVD} into Eq.~\eqref{wvd_relation_harmonic_fractional}, with $\Phi(\cdot)$ newly parameterised by $(\sigma, \theta)$, as denoted by $\Phi_{\sigma, \theta}(\cdot)$:
\begin{smallalign}
\label{wvd_relation_harmonic_fractional2}
    \Phi_{\sigma, \theta}\!\left(\omega,t\right)
        &=2\pi\!\int_{\mathbb{R}^2}
            WVD_{z}\!\left(\omega_{1},T\right)\,
            WVD_{W_{\sigma, \theta}^{(s)}}\!\left(\omega_{1},t-T\right)\,
            dT\,d\omega_{1} 
\end{smallalign}

\noindent Thus, taking the coordinate transformation $(\omega, t)\mapsto (s, t)$:
\begin{smallalign}
&\Phi_{\sigma, \theta}{\left(s, t\right)}\notag \\&= 2\pi\!\int_{\mathbb{R}^2}
            \!\!WVD_z\left(\omega_{s}(s'), T\right)
            WVD_{W_{\sigma, \theta}^{(s)}}\!\left(\omega_s(s'),t\!-\!T\right)k_{s'}
            dTds'\notag\\
    &= 2 \pi \int_{\mathbb{R}^2} \!\!WVD_z\left(\omega_{s}(s'),T\right)
\, \bar{\Pi}_{\sigma, \theta}^{(s)}(s-s',\,t-T)\,dT\,ds', 
\end{smallalign}
\noindent where $\bar{\Pi}_{\sigma, \theta}^{(s)}(s,t)$ is the base CLAWT kernel in the log-frequency domain such that:
\begin{smallalign}
   &\bar{\Pi}_{\sigma, \theta}^{(s)}(\upsilon,\,\tau) = f_0 \ln{2}\cdot 2^{\frac{s-\upsilon}{12}}\frac{e^{-\tau^{2}/\sigma_{s}^{2}}}{12\sqrt{\pi}\sigma_s}\notag\\
   &\cdot\! 
    \int_{\mathbb{R}}
      \!\exp\!\left[-\frac{\xi^{2}}{4\sigma_{s}^{2}}
          -j\omega_{s}(s - \upsilon) \xi + j\frac{\omega_{s}(s)}{\gamma}e^{\gamma \tau}\,
        2\sinh\!\left(\tfrac{\gamma\xi}{2}\right)\right]\,d\xi, 
  \end{smallalign}
\noindent and, for convenience, $\upsilon \triangleq s-s'$, $\tau \triangleq t - T$, and $\sigma_s=\sigma/\sqrt{k_s}$, as defined in Eq.~\eqref{component_wavelet}. To understand the behaviour of the kernel, $\bar{\Pi}_{\sigma, \theta}^{(s)}(\upsilon,\,\tau)$, a first-order approximation of the wavelet can be found. Firstly, note that $\frac{\omega_{s}(s)}{\gamma}e^{\gamma \tau}\, 2\sinh\!\left(\tfrac{\gamma\xi}{2}\right)\approx\omega_{s}(s)e^{\gamma \tau}\,\xi=\omega_{\text{inst}}(\tau)\,\xi$; the approximation $\sinh u\approx u$ is valid when $\gamma\xi/2$ remains sufficiently small over the principal region of the integral. Thus:
\begin{smallalign}
   &\bar{\Pi}_{\sigma, \theta}^{(s)}(\upsilon,\,\tau) \approx f_0 \ln{2}\cdot 2^{\frac{s-\upsilon}{12}}\frac{e^{-\tau^{2}/\sigma_{s}^{2}}}{12\sqrt{\pi}\sigma_s}\notag\\
   &\cdot\! 
    \int_{\mathbb{R}}
      \!\exp\!\left[-\frac{\xi^{2}}{4\sigma_{s}^{2}}
          -j\xi\left(\omega_{s}(s - \upsilon) -\omega_{\text{inst}}(\tau)
        \right)\right]\,d\xi\\
        &= \frac{f_0 \ln{2}\cdot 2^{\frac{s-\upsilon}{12}}}{6} \exp\!\left[
          -\frac{\tau^{2}}{\sigma_{s}^{2}}-\sigma_s^2\left(\omega_{s}(s - \upsilon) -\omega_{\text{inst}}(\tau)
        \right)^2\right]
  \end{smallalign}
Note also that:
\begin{smallalign}
\omega_{s}(s-\upsilon) &= 2\pi f_0\,2^{(s-\upsilon)/12},\
\\ \omega_{\mathrm{inst}}(\tau) &= 2\pi f_0\,2^{\tfrac{\tan\theta}{12}\tau+\tfrac{s}{12}},\
\\ \omega_{s}(s-\upsilon)-\omega_{\mathrm{inst}}(\tau) &= 2\pi f_0\,2^{s/12}
\left(2^{-\upsilon/12}-2^{\tfrac{\tan\theta}{12}\tau}\right)\\
&\approx -2\pi f_0 \, 2^{s/12}\left(\frac{\ln{2}}{12}\right)\left(\upsilon +\tau \cdot \tan{\theta}\right).
\end{smallalign}
In general form:
\begin{smallalign}
&\bar{\Pi}_{\sigma, \theta}^{(s)}(\upsilon,\,\tau)\notag\\&\approx \frac{f_0 \ln{2}}{6}\exp\!\left[\ln{2}\cdot\frac{s-\upsilon}{12}-\frac{\tau^{2}}{\sigma_s^{2}}
         -\sigma_s^{2}k_s^2\left(\upsilon +\tau \cdot \tan{\theta}\right)^{2}
   \right] \notag
\end{smallalign}
\noindent Given that $\sigma_s =\sigma/\sqrt{k_s}$ as per Eq.~\eqref{component_wavelet}:
\begin{smallalign}
&\bar{\Pi}_{\sigma, \theta}^{(s)}(\upsilon,\,\tau)\notag \\
   &\approx \!\frac{f_0 \ln{2}}{6}\cdot\exp\!\left[\ln{2}\cdot\frac{s-\upsilon}{12}-k_{s}\!\left(\frac{\tau^{2}}{\sigma^{2}}
         +\sigma^{2}\left(\upsilon + \tau \cdot \tan{\theta}\right)^{2}\right)
   \right]
\end{smallalign}
\noindent Thus, letting $\hat{\Pi}_{\sigma, \theta}^{(s)}(\upsilon,\,\tau)$ be the first-order approximation of $\bar{\Pi}_{\sigma, \theta}^{(s)}(\upsilon,\,\tau)$, freezing the Jacobian factor $2^{(s-\upsilon)/12}$ at $\upsilon=0$ provides:
\begin{smallalign}
&\hat{\Pi}_{\sigma, \theta}^{(s)}(\upsilon,\,\tau) \!\propto \!\frac{f_0 \ln{2}}{6}\cdot\exp\!\left[-k_{s}\!\left(\frac{\tau^{2}}{\sigma^{2}}
         +\sigma^{2}\left(\upsilon + \tau \cdot \tan{\theta}\right)^{2}\right)
   \right].
\end{smallalign}
\subsection{Training-loss definitions}
\label{sec:loss_details}
\noindent The \textit{MSE Image Reconstruction Loss}, $\mathcal{L}_{x,\text{mse}}$, is the mean squared error between the predicted and ground truth ITFRs:
\begin{smallalign*}
\mathcal{L}_{x,\text{mse}} = \frac{\lambda_{x,\text{mse}}}{NHW} \sum_{n=1}^N \|\mathbf{x}_n\! - \!\mathbf{\hat{x}}_n\|_F^2,
\end{smallalign*}
\noindent where $\mathbf{\hat{x}}_n =f_{\mathbf{\Theta}}\left(\mathbf{\Phi}_n\right)$ and $\|\mathbf{X}\|_F^2$ denotes the Frobenius norm squared:
\begin{smallalign*}
    \|\mathbf{X}\|_F^2 = \sum_{i,j} X_{i,j}^2.
\end{smallalign*}
\noindent The \textit{Log-MAE Image Reconstruction Loss}, $\mathcal{L}_{x,\text{log}}$, is the mean absolute error in the log domain, spatially weighted to focus on lower frequencies:
\begin{smallalign*} \mathcal{L}_{x,\text{log}}
    = \frac{\lambda_{x,\text{log}}}{NHW}
    \sum_{n=1}^N
    \left\|
    \mathbf{W}_x \odot
    \left[
    \log(\mathbf{x}_n+\epsilon)
    -
    \log(\mathbf{\hat{x}}_n+\epsilon)
    \right]
    \right\|_{1,1},
\end{smallalign*}
\noindent where $\mathbf{W}_x \in \mathbb{R}_{+}^{H \times W}$ is a nonnegative spatial weighting mask, with larger weights assigned to lower row indices $i$, thereby emphasising lower-frequency components. Likewise, $\epsilon$ is the \emph{sensitivity} parameter, which defines the dynamic range of the training process, and:
\begin{smallalign*}
    \|\mathbf{X}\|_{1,1} = \sum_{i,j} |X_{i,j}|.
\end{smallalign*}
\noindent The \textit{MSE Physics loss}, $\mathcal{L}_{y,\text{mse}}$, is the mean squared error between the physics prediction and the simulated physics target, across all $K$ channels:
\begin{smallalign*}
    \mathcal{L}_{y,\text{mse}} = \frac{\lambda_{y,\text{mse}}}{NKHW}\sum_{n=1}^N \sum_{k=1}^K \|\mathbf{y}_{n,k}\! - \!\mathbf{\hat{y}}_{n,k}\|_F^2.
\end{smallalign*}
\noindent For numerical stability, negative reprojection values are clipped to zero before evaluating logarithmic losses and physics-domain distributional metrics. 

The \textit{Log-MAE Physics Loss}, $\mathcal{L}_{y,\text{log}}$, is the physics error measured in the log domain with spatial weighting.
\begin{smallalign*}
    \!&\mathcal{L}_{y,\text{log}} \\
    \!&=  \!\frac{\lambda_{y,\text{log}}}{NKHW} \!\sum_{n=1}^N \sum_{k=1}^K \!\|\mathbf{W}_y \!\odot
    \!\left[\log(\mathbf{y}_{n,k}+\epsilon)\! - \!\log(\mathbf{\hat{y}}_{n,k}+\epsilon)\right]\|_{1,1},
\end{smallalign*}
\noindent where $\mathbf{W}_y \in \mathbb{R}_{+}^{H \times W}$ is a nonnegative spatial weighting mask, with once again larger weights assigned to lower-frequency components.

\noindent The \textit{Total Variation Regularisation Loss}, $\mathcal{L}_{TV}$, is the L1 norm of the spatial gradient of the predicted image, which penalises spiky artefacts:
\begin{smallalign}
    \mathcal{L}_{TV} &= \frac{\lambda_{\text{TV}}}{N} \cdot\left[\tfrac{1}{H(W-1)}\sum_{n=1}^N\sum_{i=1}^{H}\sum_{j=1}^{W-1} |\hat{x}_{n,i,j+1} - \hat{x}_{n,i,j}| \right.\notag \\
    &\left.\hspace{5mm}+ \tfrac{1}{W(H-1)}\sum_{n=1}^N\sum_{j=1}^{W}\sum_{i=1}^{H-1} |\hat{x}_{n,i+1,j} - \hat{x}_{n,i,j}|\right].
\end{smallalign}
\subsection{Component phase recovery and resynthesis}
\label{sec:phase_recovery}  

For computational efficiency, phase recovery is performed independently for each reconstructed component by estimating a single phase offset per component. Let $x[n]$ denote the input audio signal, and let $\hat{x}_k[n]$ denote the waveform associated with the $k$th reconstructed T--F component. For $n \in \mathcal{I}_k$, where $\mathcal{I}_k=\{n_{k,0},\ldots,n_{k,\mathrm{end}}\}$ denotes the temporal support of the component, the waveform is modelled as:
\begin{smallalign}
\hat{x}_k[n] &= A_k[n] \cdot \cos{\left(\phi_k[n]\right)} \\
\phi_k[n] &= \phi_{k,0} + \phi_{k,I}[n],
\end{smallalign}
\noindent where $A_k[n]$ and $\omega_k[n]$ denote the amplitude and instantaneous angular frequency of the $k$th component, respectively, as extracted from the PHAST-Net representation. The amplitudes $A_k[n]$ are obtained from the square roots of the recovered ridge-energy weights, with the inference input scaling undone. The integrated phase term is defined as $\phi_{k,I}[n] = \sum_{m=n_{k,0}}^{n-1} \omega_k[m]\Delta t$, where $n_{k,0}$ is the first detected sample of the component and $\phi_{k,0}$ is the unknown initial phase offset. The phase offset is chosen to maximise the correlation between the reconstructed component and the input signal over the component support:
\begin{smallalign}
\hat{\phi}_{k,0}
&= \operatorname*{arg\,max}_{\phi_{k,0}}
\left\{\textstyle \sum_{n \in \mathcal{I}_k} x[n] \cdot \hat{x}_k[n]\right\}
\end{smallalign}
\noindent Expanding the cosine term gives the closed-form solution:
\begin{smallalign}
    \hat{\phi}_{k,0} &= \operatorname{atan2}(-S_k,\, C_k),
\end{smallalign}
\noindent where:
\begin{smallalign}
    C_k &= \sum_{n \in \mathcal{I}_k} x[n] \cdot A_k[n] \cdot \cos{\left(\phi_{k,I}[n]\right)} \\
    S_k &= \sum_{n \in \mathcal{I}_k} x[n] \cdot A_k[n] \cdot \sin{\left(\phi_{k,I}[n]\right)}.
\end{smallalign}
\noindent The estimated offset is then used to obtain the phase-aligned waveform $\hat{x}_k[n]$ for each reconstructed component.

\begin{table}[t]
\caption{Implementations and fixed parameter settings for the quantitative benchmark.}
\label{tab:baseline_policy}
\centering
\footnotesize
\setlength{\tabcolsep}{3pt}
\vspace{-2mm}
\begin{tabular}{p{0.19\linewidth}p{0.75\linewidth}}
\hline
Method & Settings \\
\hline

Spline-PHAST-Net &
$\alpha_{\mathrm{noise}}=0.01$;
PHAST-Net followed by the fixed component-extraction and
spline-rasterisation procedure described in
Section~\ref{sec:component_extraction}. \\

SST &
Gaussian-STFT implementation;
power accumulation. \\

SET &
Gaussian-STFT implementation; same front end as SST;
IF gate $0.9$ FFT bins;
local-maximum gate;
$70$th-percentile magnitude gate;
power accumulation. \\

Reassignment &
Gaussian-STFT implementation; same front end as SST;
full two-dimensional reassignment;
threshold $10^{-10}$;
bilinear grid deposition;
power accumulation. \\

AOK &
$t_{\mathrm{lag}}=1024$,
$n_{\mathrm{rad}}=n_{\phi}=128$,
$n_{\mathrm{its}}=4$,
$\mu=0.5$,
forgetting factor $0.001$,
volume $1.0$. \\

S-Method &
Gaussian-STFT implementation; same front end as SST;
$L_f=16$;
Bartlett lag taper. \\

Choi--Williams &
Ambiguity-domain implementation;
$\gamma=0.01$ for
$\Gamma(\eta,\tau)=\exp[-(\eta\tau)^2/\gamma]$. \\

WVD &
Analytic-signal input;
linear half-sample interpolation;
full finite-record lag support. \\

RIFT &
GAFWT/LR--TV implementation;
$\alpha=10$;
temporal and frequency entropy-window standard deviations
$\sigma_t=0.5$ and $\sigma_f=10$ pixels;
$\sigma_{\ell}=1.3$;
$N=5$;
$(M(3),M(4),M(5))=(1,3,15)$;
$N_k=37$;
$K_{\mathrm{LR}}=500$;
$\lambda=2\times10^{-4}$;
$L=H=1$. \\

Spline-RIFT &
RIFT followed by IPD-assisted Kalman tracking and spline rasterisation;
peak threshold $10^{-5}$;
$\varepsilon=1$;
$\sigma_y=\Delta\omega$;
$\sigma_{\dot y}=15\Delta\omega/\Delta t$. \\

\hline
\end{tabular}

\parbox{0.96\linewidth}{\footnotesize
\vspace{1mm}
The benchmark uses a $1024\times1024$ T--F grid spanning
$100$--$1000\,\mathrm{Hz}$, with sampling rate
$f_s=11\,025\,\mathrm{Hz}$ and hop size $h=32$.
SST, SET, reassignment, and the S-Method share a common Gaussian-STFT
front end with temporal window standard deviation
$\sigma_t=0.0229\,\mathrm{s}$ and $N_{\mathrm{FFT}}=16\,384$,
selected so that the induced Cohen's-class kernel is isotropic on
the corresponding normalised T--F grid. Metrics are computed from the magnitude of each method
output. All method-specific settings were chosen before evaluation
and held constant over signals, SNR conditions, and noise realisations.
}
\vspace{-2mm}
\end{table}

\subsection{Evaluation metric definitions}
\label{sec:evaluation_metrics}

Let $R_m(t,f)\ge0$ denote the T--F representation produced by method $m$, and let $C(t,f)\ge0$ denote the reference ideal T--F representation (ITFR), generated from the model in Section~\ref{sec:itfr}. The ITFR components are rasterised using Xiaolin Wu's antialiased line algorithm \cite{wu1991aa}. Note that signed T--F representations are converted to their magnitudes. To avoid numerical sensitivity to sub-pixel discretisation, we incorporate a small soft tolerance using an isotropic Gaussian kernel $G_{\sigma_{\mathrm{eval}}}$ (with $\sigma_{\mathrm{eval}}=2$ pixels) $\widetilde{R}_m = G_{\sigma_{\mathrm{eval}}} * R_m,\,\text{and} \,\,\widetilde{C} = G_{\sigma_{\mathrm{eval}}} * C$, where $*$ denotes convolution; the same $\sigma_{\mathrm{eval}}$ is used for every method, SNR, and noise realisation. Note that ablating over $\sigma_{\mathrm{eval}}$, as reported in Table~\ref{tab:sigma_ablation} of the SM, confirms that the principal comparative conclusion is insensitive to this parameter. Let:
\begin{smallalign*}
P_m=\tfrac{\widetilde{R}_m}{\sum_{t,f}\widetilde{R}_m(t,f)},
\qquad
Q=\tfrac{\widetilde{C}}{\sum_{t,f}\widetilde{C}(t,f)}.
\end{smallalign*}
\noindent We report three complementary metrics. First, the Bhattacharyya coefficient, {\small $\mathrm{BC}(P_m,Q)=\sum_{t,f}\sqrt{P_m(t,f)Q(t,f)}$}, measures global probabilistic overlap between the method output and the reference ITFR, with larger values indicating better agreement ($0 \le \mathrm{BC}\le 1$) \cite{Bhattacharyya1943}. Second, the Jensen--Shannon divergence, {\small $\mathrm{JS}(P_m,Q)=\frac{1}{2}\mathrm{KL}(P_m\|M)+\frac{1}{2}\mathrm{KL}(Q\|M)$}, with {\small $M=\frac{1}{2}(P_m+Q)$}, measures symmetric distributional (KL) discrepancy, with lower values indicating better agreement ($0 \le \mathrm{JS} \le \ln{2}$) \cite{Lin1991}. Third, the ridge energy ratio (RER) \cite{cozens2025rift} measures the proportion of method magnitude concentrated near the soft reference ITFR ridges. Specifically, let \(d(t,f)\) denote the Euclidean pixel distance from \((t,f)\) to the union of the rasterised reference ridges. We define:
\begin{smallalign*}
W(t,f)=\exp\!\left(-\tfrac{d(t,f)^2}{2\sigma_{\mathrm{eval}}^2}\right).
\end{smallalign*}
\noindent The RER is then {\small $\mathrm{RER}(R_m,W)= \sum_{t,f}|R_m(t,f)|W(t,f)/\sum_{t,f}|R_m(t,f)|$}, where $W\in[0,1]$ is the resulting soft Gaussian ridge tube around the union of the reference ITFR ridges. This metric quantifies on-ridge concentration relative to the reference ITFR, with larger values indicating stronger localisation ($0 \le \mathrm{RER} \le 1$), and is motivated by classical notions of energy concentration and T--F localisation \cite{SlepianPollak1961I, Stankovic2001}.

\vspace{-2mm} \subsection{Supplementary  waveform equations}

\vspace{-2.85mm}{\small
\begin{equation}
\label{eq:complex_signal_log}
\begin{gathered}
x_2(t)=\sum_{i=1}^{7}\sin\!\left(
2\pi\int_{0}^{t} g_i(\tau)\,d\tau
\right),
\, 0\le t\le T,\,
T=\frac{154623}{44100}\ \mathrm{s},\\
\begin{aligned}
g_{1,2}(t)&=a_{1,2}+75t/T+75\sin(2\pi t),
& a_{1,2}&=(127.5,97.5),\\
g_3(t)&=225+75t/T,
& g_4(t)&=15+60t/T,\\
g_{5,6}(t)&=600\pm300\sin(2\pi t),
& g_7(t)&=300+600t/T .
\end{aligned}
\end{gathered}
\end{equation}
}

\vspace{-2.85mm}{\small
\begin{equation}
\label{eq:physics}
\begin{gathered}
x_3(t)=\sum_{i=1}^{4}\sin\!\left(
2\pi\int_{0}^{t} g_i(\tau)\,d\tau
\right),
\,
0\le t\le T,\,
T=\frac{154623}{44100}\ \mathrm{s},\\
\begin{aligned}
g_{1,2}(t)
    &=a_{1,2}+50t/T+50\sin(2\pi t),
& a_{1,2}&=(85,65),\\
g_3(t)
    &=150+50t/T,
&
g_4(t)
    &=10+40t/T.
\end{aligned}
\end{gathered}
\end{equation}
}

\end{document}